\documentclass[twoside,twocolumn,9pt]{revtex4}
\usepackage{graphicx} 
\usepackage[greek,english]{babel}
\usepackage[T1]{fontenc}
\usepackage[usenames,dvipsnames]{xcolor}

\usepackage{dcolumn}
\usepackage{color}
\usepackage{ulem}
\usepackage{mathptmx, helvet, courier}

\newcommand{\rout}{\bgroup\markoverwith%
{\textcolor{red}{\rule[.5ex]{2pt}{0.5pt}}}\ULon}

\newcommand{\gout}{\bgroup\markoverwith%
{\textcolor{purple}{\rule[.5ex]{2pt}{2pt}}}\ULon}

\newcommand{\gb}{\greektext{}b\latintext{}}
\newcommand{\ga}{\greektext{}a\latintext{}}

\usepackage{xr}
\usepackage{epstopdf}

\definecolor{cream}{RGB}{222,217,201}

\begin{document}

\title{Unfolding knots by
    proteasome-like systems:
simulations of the behaviour of folded and neurotoxic proteins}

%

\author{Micha{\l} Wojciechowski}
\thanks{These authors contributed equally to this work.}
\affiliation{Institute of Physics, Polish Academy of Sciences, Al. Lotnik\'ow 32/46,
PL-02668 Warsaw, Poland}
\author{\`{A}ngel G\'{o}mez-Sicilia}
\thanks{These authors contributed equally to this work.}
\affiliation{Instituto Cajal, Consejo Superior de Investigaciones Cient\'ificas, (CSIC),
Av. Doctor Arce, 37, 28002 Madrid, Spain}
\affiliation{Instituto Madrile\~no de Estudios Avanzados en Nanociencia (IMDEA-Nanociencia),
C/ Faraday 9, 28049 Cantoblanco, Madrid, Spain}
\author{Mariano Carri\'{o}n-V\'{a}zquez}
\affiliation{Instituto Cajal, Consejo Superior de Investigaciones Cient\'ificas, (CSIC),
Av. Doctor Arce, 37, 28002 Madrid, Spain}
\affiliation{Instituto Madrile\~no de Estudios Avanzados en Nanociencia (IMDEA-Nanociencia),
C/ Faraday 9, 28049 Cantoblanco, Madrid, Spain}
\author{Marek Cieplak}
\email{mc@ifpan.edu.pl}
\affiliation{Institute of Physics, Polish Academy of Sciences, Al. Lotnik\'ow 32/46,
PL-02668 Warsaw, Poland}

%
%
\begin{abstract}
Knots in proteins have been proposed to resist proteasomal
    degradation. Ample evidence associates proteasomal degradation with
    neurodegeneration. One interesting
    possibility is that indeed knotted conformers stall this
    machinery leading to toxicity. However, although the proteasome is known to
    unfold mechanically its substrates, at present there are no experimental
methods to emulate this particular traction geometry. Here, we 
consider several dynamical models of the proteasome in which the complex is
represented by an effective potential with an added pulling force.
This force is meant to induce translocation of a protein
or a polypeptide into the catalytic chamber. The force is either constant
or applied periodically. The translocated proteins
are modelled in a coarse-grained fashion. We do comparative analysis of
several knotted globular proteins and the
transiently knotted polyglutamine tracts of length 60
alone and fused in exon 1 of the huntingtin protein. Huntingtin is associated with
Huntington disease, a well-known genetically-determined neurodegenerative disease.
We show that the presence of a knot
hinders and sometimes  even jams translocation.
We demonstrate that the probability to do so depends on the
protein, the model of the proteasome, 
the magnitude of the pulling force, and the choice of the pulled terminus.
In any case, the net effect would be a hindrance in the proteasomal degradation
process in the cell. This would then yield toxicity \textit{via} two different
mechanisms: one through toxic monomers compromising degradation and another by
the formation of toxic oligomers.
Our work paves the
    way to the mechanistic investigation of the mechanical unfolding of knotted
structures by the proteasome and its relation to toxicity and disease.
\end{abstract}

\maketitle



%

%
%
%
%
%
%
%

\section*{Introduction}
Neurodegenerative diseases are poorly understood and
represent one of the major challenges to modern medicine~\cite{diseases_of_the_old}. Among these maladies, many are known to be tightly
related to proteins that are present in the human brain. These proteins show
similar traits. One is the tendency to 
form amyloid fibers
within neurons and/or 
outside them~\cite{cross_beta_spine,a11}.
Another is the high mechanical polymorphism as assessed by single-molecule
force spectroscopy~\cite{neurotos} for polyglutamine tracts
(polyQ, also denoted as Q$_n$, where $n$ is the number of residues) 
as well as \ga{}-synuclein, \gb{}-amyloid, 
and Sup35NM. All of them are amyloidogenic 
intrinsically disordered proteins (IDPs)~\cite{neurotos}.
In equilibrium, IDPs may adopt a number of different conformations and
interconvert 
in a nanosecond timescale~\cite{ns_fluctuations}.

The mechanical polymorphism of Q$_n$ has been demonstrated to follow
from a conformational one,  which was characterized theoretically for
$n$ between 16 and 80 in ref.~\cite{Angel} by following the 
meta-dynamics methods used by Cossio {\it et al.} for 
polyvaline (polyV, denoted also as V$_n$)~\cite{val60}.
In particular, it has been found that about
9.3\% of the 246 structurally independent Q$_n$ conformers obtained for 
$n=60$ were knotted compared to about 3.6\% of the structures obtained for
polyV of the same length~\cite{Angel}. 

It should be noted that some knotted structures can be spotted
in the CATH database~\cite{cath} which reperesents
superfamilies of proteins
as classified by their secondary structure motifs.
We have checked that the previous version of CATH used for making
comparisons in refs.~\cite{Angel} and~\cite{val60} (5403 structures with
the sequential length smaller than 250) contains 71 knotted proteins, smaller than 0.5\%. 
The KnotProt database~\cite{Sulkowskabaza}, devoted to proteins with knots,
indicates that the smallest known knotted protein is  MJ0366
from {\it Methanocaldococcus jannaschii}~\cite{knots2} -- it
comprises 87 residues, which is above the range of length considered in the
Q$_n$ systems in~\cite{Angel}. Its folding pathways have been studied
in refs.~\cite{Micheletti,shallowJoanna,knotshallow}. 

One of the knots
in the Q$_{60}$ set~\cite{Angel} was of the three-twist ($5_2$) kind, with five
intersections. Such a knot can be found in the Protein Data Bank (PDB)~\cite{pdb}.
In particular, it is present in ubiquitin hydrolase UCH-L3 (PDB code 1XD3), 
a protein that deconjugates ubiquitin before degradation in the proteasome.
Thus, the knot in this protein has been hypothesized to confer
resistance to proteasomal unfolding~\cite{knots2}.
All of the remaining knots identified in
Q$_{60}$ and all the ones found in V$_{60}$ were of the common trefoil kind ($3_1$).
No knots were found for $n$ smaller than 35. Since the polyQ systems are
IDPs, all the conformations, including the knotted ones, are transient.
However, knots in polyQ last for hundreds of
    nanoseconds~\cite{Angel}, two orders of magnitude longer than unknotted
conformations.
Properties of the globular knotted proteins are reviewed in refs.~\cite{VirnauJackson,Sulkowska,LimJackson}.

V$_{60}$ is an artificial protein, in the sense that it cannot
be found in nature. It was constructed
to demonstrate that only a fraction of possible folds is adopted by proteins
collected in the CATH database, implying a selective role of evolution,
and perhaps also of physical constraints,
in favouring certain folds~\cite{val60}.
This topic has been discussed for IDPs in ref.~\cite{Konrat}.
However, long polyQ tracts  arise in nature near the N-terminus
of several proteins, such as huntingtin,
Atrophin-1 (ATN1), androgen receptor (AR) and several ataxins (ATXN). 
Huntingtin is known to
be essential in embryonic development~\cite{htt_development}, and is considered
to be related to gene expression regulation~\cite{htt_gene_expression}
as well as to anchoring or transport of vesicles~\cite{htt_vessicles}.

Nonetheless, before translation, the genes encoding for these proteins contain
long cytosine-adenine-guanine (CAG) repeats, which encode for the polyQ region. 
These CAG repeats are prone to  
due to slippage of the DNA polymerase~\cite{polymerase_slippage} that elongates the
number of CAG repeats in the gene and, therefore, the length of the polyQ tracts.
If the length of the polyQ tract in the aforementioned proteins is greater than a
(disease-dependent) threshold, they cause neurodegenerative
diseases such as Huntington,  dentatorubropallidoluysian atrophy (ATN1),
spinal and bulbar muscular atrophy (AR) and many spinocerebellar ataxias (ATXN).
Huntington disease, in particular, leads to progressive motor and cognitive
impairments. The non-pathological length of the polyQ tract in 
huntingtin may start at
7 and is usually between 16 and 20 repeats~\cite{Warby}, while the pathogenic
threshold is around 35 residues.

The mutant huntingtin protein easily forms
toxic oligomers and highly ordered amyloid fibers~\cite{DiFiglia}
and the toxicity of the olygomers may result from
 interactions with the membrane and perturbation of the calcium
regulation~\cite{toxic_oligomers1,toxic_oligomers2}. Nonetheless, other 
mechanisms point to toxicity appearing at different stages: either in the 
fibrillar state or at the monomeric level.
In particular, a cell microinjection assay~\cite{nagai_toxic_monomer}
has proved the toxicity of the monomers.
The mechnisms of the monomeric toxicity can be related to the damage of the
degradation machinery~\cite{neurotos,impairment_proteasome}. Here, we
propose  that the toxicity may be due
to the presence of knots in Q$_n$ with $n$ exceeding 
the disease threshold of about 35 -- the
typical sequential size, $\Delta k$, of a knot found in Q$_{60}$~\cite{Angel}.

The knot-end locations are defined operationally
through systematic cutting-away of the residues from both termini until 
the knot disintegrates~\cite{km1,taylor}. Location $k_-$ denotes the end that
is closer to the N-terminus and $k_+$ the one further away.  The knots in
Q$_{60}$ are shallow since $k_+$ is always near the C-terminus, between
sites 53 and 58, whereas $k_-$ is between 16 and 23 \cite{Angel}.
Nonetheless, since the polyQ tract is close to
the N-terminus of huntingtin, 
a knot in it would actually be deep.
The evident lack of the N-to-C-termini symmetry reflects 
most likely the directionality of
the peptide chain. 
Nevertheless
the locations of the Q$_{60}$ knot ends are seen to be much better defined
and closer apart than those of the V$_{60}$ ends~\cite{Angel}. 

For the sake of comparison with Q$_{60}$, we also consider four
proteins that have a (non-transient) knot in their native state
as they provide a well defined reference situation.
Three of them are deeply knotted: the frequently studied methyltransferase
YibK (PDB code 1J85 chain A, 156 residues)~\cite{knots2,VirnauJackson,chwastdeep},
RNA methyltransferase YBEA~\cite{LimJackson} (1NS5 chain A, 153 residues),
and ubiquitin hydrolase UCH-L3~\cite{knots2} (1XD3 chain A, 229 residues). 
The fourth one, the just mentioned MJ0366~\cite{LimJackson}
(2EFV chain A, 82 residues), has a knot that is shallow. 
UCH-L3 contains a three-twist  knot and the remaining proteins
have trefoil knots. We shall use the PDB structure codes to refer
to these proteins in a unified manner.

The relevance of the conformations with knots is that they may derail
the degradation processes in the recycling machinery of the cell
that are carried out by
proteasomes, which would then lead to an
increase in the concentration of the neurotoxic 
conformers. Derailing
of the degradation machinery may occur either through 
a significant elongation of the degradation time
or by an abandonment of the process only after a
partial proteolysis.
The idea that knotted globular proteins may disable
proteasomes has been suggested by Virnau {\it et al.}~\cite{knots2}
in the context of protein UCH-L3 but it has not been demonstrated. Here,
we use a simplified model of the proteasome~\cite{Wojciechowski}
and show that indeed jamming may take place both for the globular
knotted proteins and for the transient knotted conformations
in IDPs.
We show that the degree  to which the degradation is disabled
depends on the protein, on its conformation, and on the
terminus at which the intake into the proteasome starts.
It also depends on the value of the effective pulling force of the
molecular motor involved: the lower the force, the bigger
the impediment. 
Whenever the probability of derailing the process of degradation is
non-zero, an accumulation and aggregation of the toxic molecules 
would set in.

\begin{figure}
\begin{center}
\includegraphics{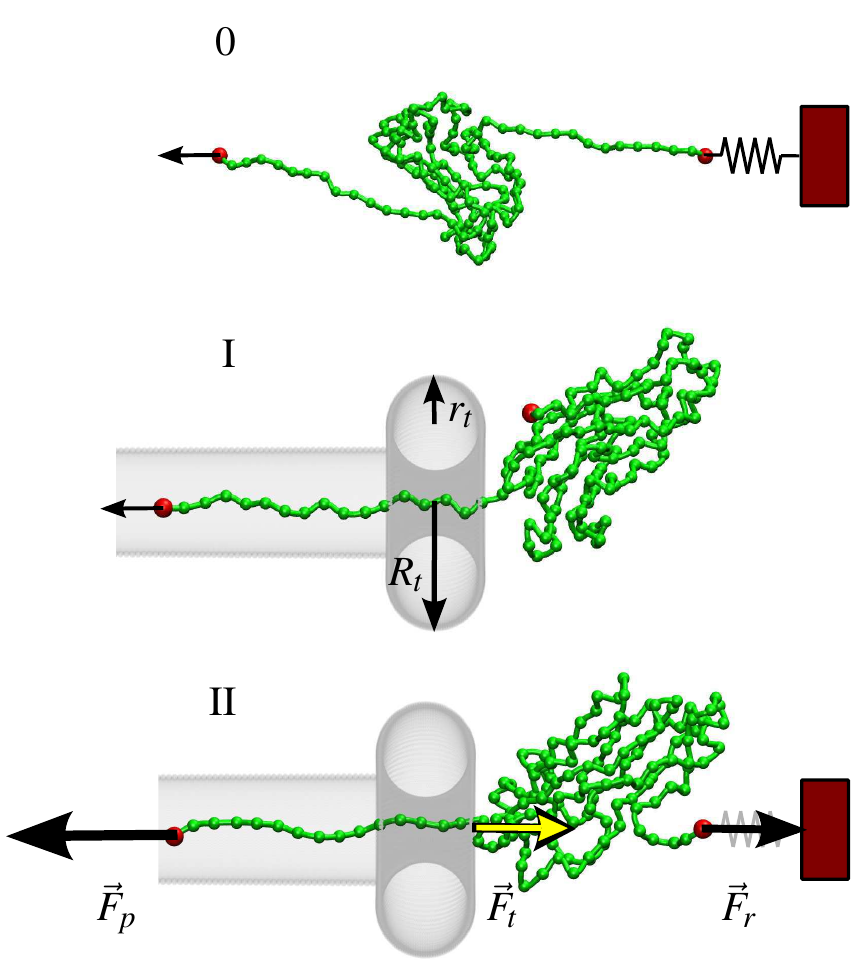}
\caption{\label{situation} 
Protocols of pulling considered in the paper. Protocol 0 is the
typical AFM-like pulling. Protocol I corresponds to
the natural action of the proteasome. Protocol II is used in the
measurements of the stalling force. The bottom panel
defines the forces involved, as discussed in the text.
}
\end{center}
\end{figure}

Although the term ``proteasome'' applies to  the 
protein-degrading complexes
in eukaryotes and archea~\cite{Coux_1996}, we use it here 
in a general sense. In bacteria, examples of 
similar complexes are ClpXP and Lon 
proteases~\cite{Goldberg_1990,Gottesman_1996}.
The process of protein degradation is carried out in 
ATP-dependent proteases. All such proteses share a common shape: four
stacked rings forming an axial channel and an inner
catalytic chamber.
We model the proteasome complex to be funnel-shaped and be endowed with
impenetrable walls~\cite{Wojciechowski}. The funnel is composed 
of a torus that is fused together  with a long cylinder  as shown
in the middle panel of Fig.~\ref{situation}.
The detailed value of the geometrical parameters in our model
are meant to render the structure of the
proteasome derived in refs.~\cite{Zhang_2009_struct,Groll_1997}.

A proteasome operates as a molecular engine in that it attracts a
(tagged) protein into its entrance and translocates it down its channel
into the catalytic chamber where the protein is lysed into short peptides. 
In our model, the mechanical aspects of this action are mimicked
by an application of a pulling force $F_p$ whereas degradation itself
is identified with translocation through a reference plane.
An alternative and more detailed way to imitate
the mechanics involved is to impose allosteric transformations in the
proteins constituting the proteasome~\cite{Stan,Tonddast,Kravats-Tonddast}.
However, our simpler model 
has the advantage of accessing long time scales
and generating a substantial number of trajectories that are necessary 
to statistically assess the role of the knotted structures. It also allows for
a straightforward comparison with the situations in which the force
is applied periodically, the pulling is performed at a constant speed,
$v_p$, or when  mechanical unfolding takes place 
without the proteasome. This simplified model does not describe
the actual proteolysis nor the release of partially degraded proteins.

We have used this model~\cite{Wojciechowski} to characterize the nature of protein
unfolding in the proteasome and to demonstrate that the stalling force 
measured by single-molecule force spectroscopy techniques~\cite{Bustamante,Aubin-Tam_2011}
may be smaller than the traction force of the unfoldase motor
because one needs to include a mechanical reaction force, $F_t$, 
of the proteasome against which the protein is pushing, as shown
schematically in Fig.~\ref{situation} (bottom panel).
Our model uses a simplified coarse-grained model
of the protein~\cite{Hoang,Hoang2,JPCM,models,plos} in which the amino
acid residues are represented by effective atoms located at the positions
of the $\alpha$-C atoms and the solvent is implicit. 
The characteristic value of $F_p$ is, as of yet, unknown. Nonetheless, it has
been demonstrated~\cite{Bustamante} that degradation takes place at
a rate of about 80 residues per second. Our approach in ref.~\cite{Wojciechowski}
was  to determine the characteristic time of translocation
as a function of $F_p$ and then to extrapolate it to the time
corresponding to the degradation of the studied protein at the
experimental rate. Here, we study properties as a function of $F_p$
to get the sense of changes involved but do not actually extrapolate.

We consider three protocols of protein unfolding as illustrated 
in Fig.~\ref{situation}. Protocol 0 involves no proteasome:
one terminus is attached to an anchored elastic spring while the
other one is pulled. In protocol I, one terminus is dragged
into the proteasome and the other is free. Protocol II is similar,
but the previously free terminus is attached to a spring which exerts a 
backward force, $F_r$. 
Pulling may be applied either to the N- or the C-terminus. 
The specific protocol used
will be denoted as, for example, I-N and I-C respectively.
It must be noted that 
obstruction for degradation in the  I-N protocol
 does not imply an obstruction in the I-C
protocol. The obstruction shows as jamming in our model
and we shall use this term as a shorthand designation.
In each of these protocols, pulling may take place in various modes.
We consider the cases of a constant pulling speed, a constant pulling
force and a force applied periodically. The last mode is the
closest to how the proteasome actually operates since the process is
controlled by the intermittent supply of the ATP molecules.
Measurement of the stalling force has been done
using mode II in 
experiments~\cite{Bustamante,Aubin-Tam_2011}.

Stalling occurs when the protein is unable to translocate. In the
experiments performed in refs. \cite{Bustamante,Aubin-Tam_2011}, this happened  when 
the opposing force $F_r$ is such that there 
is no translocation. We have argued~\cite{Wojciechowski} that $F_r$ is
not equal to $F_p$ as commonly assumed, but to $F_p - F_t$ by the
third law of mechanics.
Here, we revise this issue and show that using
an all-atom model of the protein combined with an explicit
solvent also supports our previous finding about the interpretation 
of the measured stalling force.
It should be noted that in our model, we separate all forces
acting on the protein into two parts: the motor-like pulling 
and the reaction of the proteasome that resists the entry because
of steric interactions. Our interpretation of the stalling experiments
experiments is that one measures the total
force $F_p-F_t$ and not just $F_p$.

\section*{Materials and Methods}

Our model of the proteasome has been introduced 
recently in ref.~\cite{Wojciechowski}. It 
assumes that the funnel-like geometry of the proteasome complex can be represented
as an effective external potential, in analogy to several models
of translocation through cylindrical channels~\cite{Muthukumar,Makarov_2004,Makarov_2005,West2006,Makarov_2009,Szymczaktrans,Tian2005}.
The potential has axial symmetry, which reflects the fact
that the two substructures of the biological proteasome -- the 20S core
particle, made out of four heptameric rings, and the two 19S regulatory caps 
-- share an axial channel. The outer two rings
of 19S caps form a gate where the polyubiquitin chains
that are attached to the protein get recognized so that the protein
is directed to the core chamber, leaving the chains behind
(in the case of ClpXP, the proteins are tagged by short peptides at a 
terminus and the tags enter the proteolytic chamber).

The catalytic action takes place in the inner
surface of the two rings in the middle of 20S. 
The target protein enters one of the caps and then moves
further down the axial channel. The cap is represented by a
torus and the channel by a cylinder. The channel is considered to be
indefinitely long as we do not model the actual degradation. As
a result, only one cap is included in the model and the protein
cannot emerge ``on the other side'' and refold.
The combined torus-cylinder shape defines the funnel.

The radius of the channel has been found experimentally to be
$\approx 7.5$~{\AA}~\cite{Zhang_2009_struct,Groll_1997}. It is defined by the average
distance between opposing heavy atoms on the inner side of the proteasome.
We consider the channel to have a radius $r_c=8$~{\AA} to account for
flexibile adjustments of the channel. 
This channel is wide enough to accomodate a hydrated polypeptide
and some elements of secondary structure.
The torus is described by the equation
$(x^2 + y^2 + z^2 + R_t^2 - r_t^2)^2 \;=\; 4 R_t^2(x^2+y^2)  $,
where the major radius is $R_t=13$~{\AA}
and the minor radius is $r_t=6$~{\AA}.
The opening in the narrowest place of the torus has a radius of 
7~{\AA}, which is smaller than the radius of the cylindrical channel
describing the core particle.
This disparity accounts for
an extra cavity-like space that forms between the cap and the core particle.

In protocols I-C and I-N, the reference plane is defined as one shifted
25~{\AA} inward with respect to the equatorial plane of the torus -- where the opening
in the funnel is the narrowest. 
The shift is introduced to make sure that the rear terminus is
definitely within the cylinder.
In protocol II, the criterion
involved is arriving at an end-to-end extension which is
equal to 85\% of the backbone length. If a protein is knotted at the beginning, the
evolution of the knot is followed during unfolding to check whether at
the end it has tightened, untied or remained in place.

The interaction between the effective atoms of the protein
and the surface of the torus as well as the inner
surface of the cylindrical channel is assumed to be repulsive and given by the
truncated Lennard-Jones (LJ) potential as follows
\begin{equation}\label{eq:LJcut}
V^s_{i}  = \left\{ \begin{array}{lll}
 4\epsilon \left[ \left(\frac{\sigma}{d_{i}} \right)^{12} - \left(\frac{\sigma}{d_{i}} \right)^6 \right]&, & d_{i} \le r_{min} \\
0&,       & d_{i} > r_{min}
 \end{array} \right.
\end{equation}
where $d_i$ is the smallest distance between the $i$th residue  and the torus/channel
surface. The distance $r_{min}=6$~{\AA} is the position of the minimum of
the potential, which takes into account excluded volumes of residues 
and wall atoms. We take
$\sigma\;=\;0.5^{1/6}r_{min} = 5.345$~{\AA}.

The energy parameter $\epsilon$ comes from our coarse-grained 
struc\-ture-based~\cite{Go0,Takada} mod\-el of the sub\-strate
pro\-tein/pep\-tide~\cite{Hoang,Hoang2,JPCM,models,plos}
--  it is e\-qual to the depth of the LJ po\-ten\-tial as\-so\-ci\-at\-ed
with the  con\-tacts:
$V_{ij}(r) = 4\epsilon \left[ \left(\sigma_{ij}/r_{ij} \right)^{12} -
\left(\sigma_{ij}/r_{ij} \right)^6 \right]$.
The con\-tacts are de\-ter\-mined by us\-ing
the criterion of atomic overlaps -- the procedure denoted as 
OV in the detailed description of several approaches to
the contact-map determination presented in ref.~\cite{Wolek}.
These contacts are only native in the case of globular proteins, whereas
for the polyQ chains they are determined from the conformation
at hand~\cite{Angel}. The value of the $\epsilon$ is approximately
equal to 106~pN$\cdot${\AA}, which correlates well with the 
experimental data on stretching at constant speed, after extrapolating
to the speeds used in the experiments~\cite{plos,Wolek,Current}.
The length parameters $\sigma_{ij}$ in the contact potentials
are determined so that the potential minimum agrees with the native distances
between $i$ and $j$.

Bonded interactions are modelled by a harmonic potential
with the spring constant of 50~$\epsilon/${\AA}. 
The local backbone stiffness is described by the chirality
potential~\cite{models}. The solvent is represented by damping
and random fluctuational forces, the amplitude of which depends on the
temperature, $T$. Our simulations are performed at $k_BT=0.3~\epsilon$
which is close to the optimal folding $T$ of the model proteins~\cite{Hoang2}
 ($k_B$ is the Boltzmann constant). With our calibration of $\epsilon$
this choice is also consistent with studies performed in a
vicinity of the room temperature. The time unit, $\tau$, in the simulations
is of order 1~ns due to overdamping.

The initial placement of the protein/polypeptide relative to the
proteasome model is based on the following procedure.
Both termini are located on the main axis of the
proteasome and away from the torus.
The protein is then
moved towards the torus in small steps until one of the
heavy atoms collides with it.
The placement just before this last step is taken as the starting
state. Henceforth, in the coarse-grained model, only the $\alpha$-C 
representation  is used dynamically.
The coordinate reference system is chosen so that the central axis 
of the cylinder is along the $z$ axis.
In protocol I, translocation is said to take place if all of the
system moves past the reference plane defined before.
The time needed to accomplish this defines the translocation time, $t_{T}$.
In protocol II, the characteristic time corresponds to 
achieving a nearly full extension (85\%).

In the part where polyQ is considered, the structures are taken
from~\cite{Angel}. HTT, which stands here for the exon 1 of huntingtin,
was obtained by homology modelling using MODELLER~\cite{modeller}.
Knotted and unknotted HTT are based on two templates: The structured
regions were taken from the X-Ray-resolved 17-glutamine structure under the
PDB code 3IOR~\cite{htt_Q17}, while the (knotted or unknotted) polyQ from
ref.~\cite{Angel} were used as templates for the polyQ region. 10 models
were done for each conformation, each one was checked to preserve the
knotting state (HTT must be knotted if polyQ was knotted and \textit{vice
versa}) and the lowest-energy configuration from those was chosen to
stretch.

In the final part of this work,
we consider an unknotted protein which represents
a model protein in the field of single-molecule force spectroscopy,
the I27 domain of cardiac titin whose PDB
structure code is 1TIT
and which has a net charge of $-6~e$. We reexamine
the issue of the stalling force -- now by using all-atom simulations.
These simulations were performed by using the GROMACS 4.6.5 package~\cite{gromacs} with the AMBER99 force field~\cite{amber}. The
molecules of water are described within the TIP3P model~\cite{TIP3P}.
The time integration of the equations of motion was performed using 
the leap-frog algorithm with a time step of 1~fs.
We have added 64 Na$^+$ and 58 Cl$^-$ ions to neutralize the system
and to keep the nearly physiological ionic strength of  150~mM.
The GROMACS-based model was augmented with the potential due to the
proteasome walls (eq.~\ref{eq:LJcut}) constructed in the same fashion 
as in the coarse-grained simulations  except that now the cylinder
needs to have a finite length. The system was
placed within a cuboid box  with the extension of 100~{\AA}
in the $x$ and $y$ directions and  200~{\AA} in the $z$ direction.
The long direction of the cuboid coincided with the central axis of the
 proteasome model and the equatorial plane of the torus is 135~{\AA}
away from the bottom the system.
Periodic boundary conditions were used.

The water molecules are allowed to be above the plane corresponding
to the top of the torus and within the inside of the  proteasome model.
Due to the periodic boundary conditions, the water molecules can also
be at the bottom of the cuboid box, {\it i. e.} underneath the end
of the cylinder (which has a finite length in all-atom simulations).
There are also two repulsive walls so that the cylinder is not
surrounded by water. One is at the bottom of the cylinder (at 7~{\AA} above the
bottom of the whole system) and another at the top of the torus (away from the
center).
The equilibration of water in the initial stage was implemented through 
1~ns while holding the protein frozen in its native state.

When determining the forces of interactions with the proteasome walls, all the
atoms of the protein and water are considered as having a radius of 2~{\AA}
and the wall itself is thought of as being made of atoms also
with a radius of 2~{\AA}.
A purely repulsive potential for interactions with the wall of the funnel
leads to an expulsion of water from the  proteasome model (without
affecting the protein). In order to
hold the water in, we consider a modified attractive LJ potential
-- augmented by the term $A\;r +B$ -- where the parameters $A$ and $B$ 
were selected so that both the force and the potential vanish at 7~{\AA}
(slightly smaller than the radius of the cylinder).
The energy parameter was taken to be 4~kJ/mol.
It should be noted that this value is an order of magnitude
larger than, say, the C--C energy parameter in the AMBER99 force field,
0.359824~kJ/mol. The reason for this resides in the fact that the cylinder has no atomic structure
and larger values are needed. VMD~\cite{vmd} has been used for the
representation of protein structures.

\section*{Results}

\subsection*{Constant velocity}

There is no constant speed pulling in real
proteasomes. However, this mode of action allows one
to derive a characteristic force which is also of relevance
for constant force situations as it approximately corresponds to
a crossover between the low and high force behaviors~\cite{clamp}.
Consideration of the constant speed processes also sheds light on possible 
hindrances to translocation and allows for a direct comparison to 
typical constant-velocity experiments in atomic force microscopy (AFM).

At constant $v_p$, one monitors the tension,
$F$, in the backbone and studies it
as a function of the displacement, $d$, of the spring that is
attached to the pulled end of the protein.
The characteristic force, $F_{max}$, associated with a protein is defined
as the height of the maximal isolated force peak. It depends on $v_p$ in a 
weak way (approximately logaritmically) so one may state that $F_{max}$
refers to a certain narrow range of relevant forces. We take
$v_p$ to be 0.005~{\AA}/$\tau$, which is typically two orders of magnitude
faster than in the AFM experiments (in the absence of a proteasome).

In addition to determining $F(d)$, we monitor the time dependence of the
sequential location of the knot ends. When pulled without the proteasome,
the knot ends have been discovered to jump to preferred 
locations~\cite{knotjumps},
instead of moving in a diffusive manner, until getting tightened maximally.

\begin{figure}
\begin{center}
\includegraphics{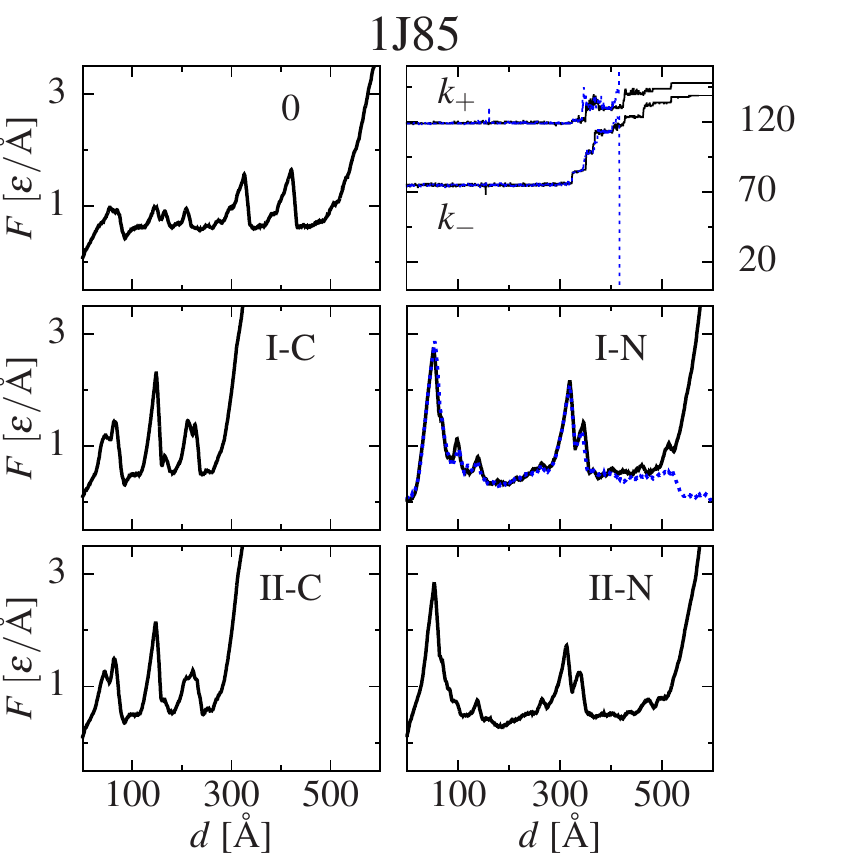}
\caption{\label{vel1J85} 
Example $F$--$d$ recordings for 1J85 in the different pulling protocols at a constant speed.
The top-right panel shows the locations of the knot ends 
in protocol I-N for two trajectories.
In one trajectory (dotted blue lines), the knot unties and 
the protein translocates. In the other (solid black lines),
the protein jams the channel and the tightened knot arrives at a permanent location.
The profiles have been selected out of 100 trajectories
for each protocol.
The profiles shown in the middle-right panel correspond to the
trajectories shown in the top-right panel.
 }
\end{center}
\end{figure}

\begin{figure}
\begin{center}
    \includegraphics{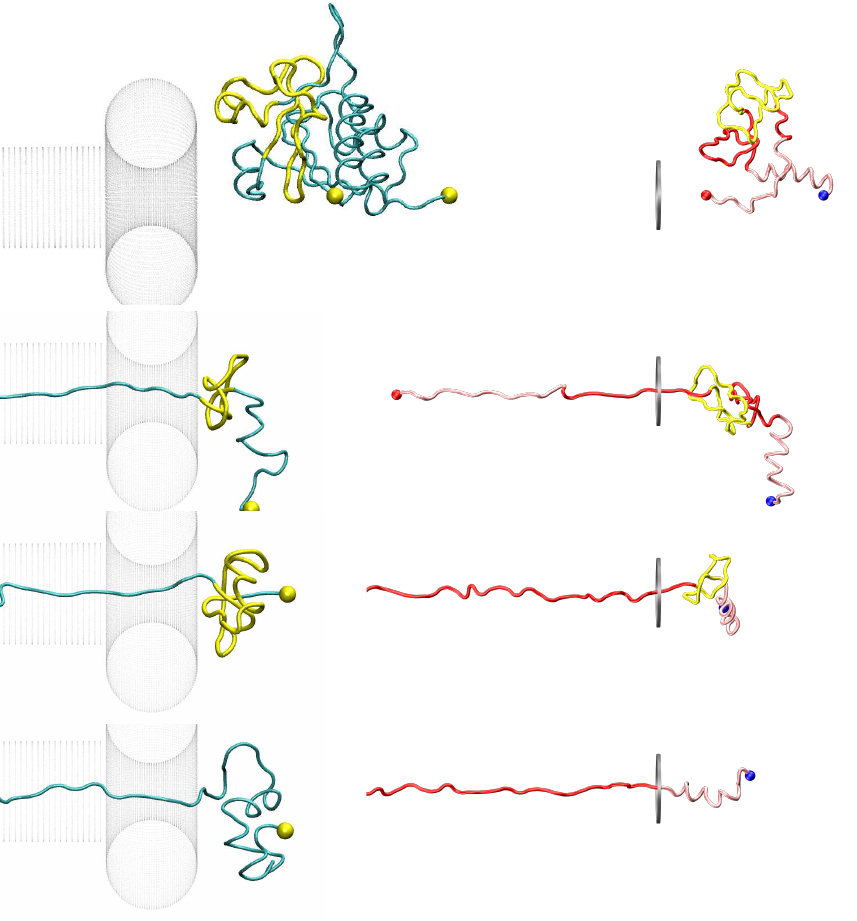}
\caption{\label{stuckvel} 
    Two example trajectories that lead to
    translocation of knotted species. The left panels show translocation  of 1J85 
    when pulled in the I-N protocol and the right ones the unfolding of
    a HTT$_{60}$ knotted conformation. The knot is marked with yellow. In the
case of HTT$_{60}$, the polyQ part is marked in red.
 }
\end{center}
\end{figure}

Fig.~\ref{vel1J85} provides an illustration of the $F(d)$ plots for 1J85
obtained in the five protocols of pulling. We have considered 100 trajectories.
For the I-N protocol about 50\% of the trajectories lead to jamming
combined with the tightening of the knot.
The motion of the knot ends along the sequence is
shown in the top-right panel of Fig.~\ref{vel1J85}.
The remaining 50\% lead to translocation, as shown in the left panels of Fig.~\ref{stuckvel},
combined with untying of the knots in which the knot ends
move to the termini, which means that the knot disappears.
In all other protocols, all trajectories lead to jamming in 
which the knot may or may not change location. Typically, jamming
involves the knot being unable to enter the cylindrical
channel.

The $F(d)$ patterns are seen to be visibly affected by the
presence of the proteasome and, in particular, the values
of $F_{max}$ are generally different than those in protocol 0.
Note that stretching in protocol 0 invariably leads to knot
tightening whereas the action of the  proteasome model may also lead
to the knot slipping off the backbone, especially if the knot is shallow.
It is interesting to note that knotted proteins pulled by the proteasome have 
$F_{max}$ larger than in its absence.
On averaging over the trajectories, for 1J85 we get $\langle F_{max}\rangle$ of
1.85, 2.67, 3.13, 2.48, and 3.09~$\epsilon/${\AA}
for  protocols 0, I-C, I-N, II-C, and II-N respectively.
For 1NS5, the corresponding numbers are 1.97, 2.71, 2.50, 2.42
and 2.69~$\epsilon/${\AA}.

In the case of 1NS5 (see S1~Fig.),
protocols I-C and II-C may lead either to translocation or
to jamming whereas protocol II-N leads only to jamming. For protocol I-N,
only about 4\% of the trajectories end up in translocation.
The knot is located near the C-terminus so one would expect that
it unties readily during pulling by the N-terminus;
however, the opposite is observed.
In the case of 1XD3 (see S2~Fig.), almost all I-C and II-C
trajectories lead to translocation. For I-N, 35\% trajectories
unfold and for II-N -- none. The values of
$\langle F_{max}\rangle$ exceed or equal (in the case of protocol II-N)
the $\langle F_{max}\rangle$ for protocol 0. For the shallowly knotted
2EFV (see S3~Fig.), $\langle F_{max}\rangle$
in protocols I and II is close to that in protocol 0 (within
0.1~$\epsilon/${\AA}). The knot easily unties in protocols
I-N and I-C and translocation is unhindered. However, in
protocol II-C, there is no instance of translocation whereas
in II-N translocation occurs with a probability of 30\%.
The knot ends are at sites 11 and 73 (the full length of the
protein is 87, but the structure of the first five residues
is unknown and therefore the model does not include them).

We conclude that keeping the backward terminus supported, as in the
experiments on the stalling force, affects the physical
outcome of the process. For instance protein 2EFV translocates
in protocol I-C but it does not in protocol II-C.
Also, protocol II precludes untying of knots.

\begin{figure*}
    \centering
    \includegraphics{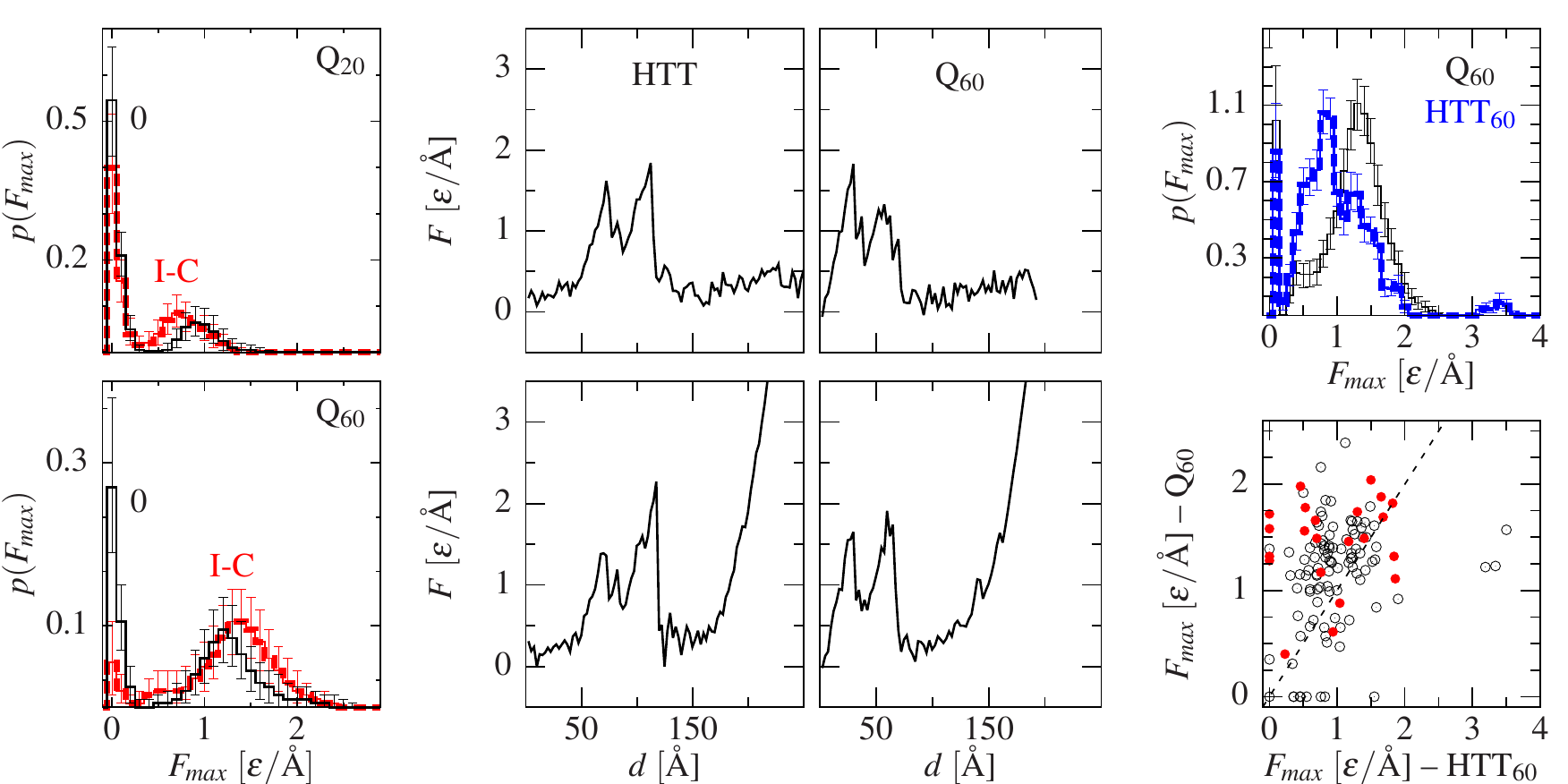}
    \caption{\label{polyq_cv} Constant velocity pulling analysis of polyQ. The
        left panels show a distribution of mechanical stabilities for the
        Q$_{20}$ and Q$_{60}$ conformations when pulled in protocol 0 or I-C.
        The two middle panels show examples of F--d recordings of the
        proteasomal unfolding
        --protocol I-C-- of knotted HTT (left) and Q$_{60}$ (right)
        molecules that translocate (top) or induce jamming
        (bottom). All panels correspond to the same starting conformation, thus
        the similarity in the unfolding pattern.
        The right panels compare the mechanical stability of 
        Q$_{60}$ with and without the exon 1
        handles of huntingtin. The former is called HTT$_{60}$.
        The top panel shows the distribution
        of mechanical stabilities, while the bottom panel shows a scatter plot
        comparing the mechanical stabilities in each of the
        cases, with the knotted conformations highlighted in red.
        In this analysis, $F_{max}=0$ means there is no articulated force peak
        greater than the thermal noise ($0.1~\varepsilon/$\AA{}) in
        the force--extension plot. The error bars in the histograms represent 
        a 95\% confidence interval.
    }
\end{figure*}

We now consider the constant velocity pulling of Q$_{20}$ and Q$_{60}$,
below and above the pathological threshold, respectively,
of most polyQ-related diseases including Huntington.
Using the conformations obtained in ref.~\cite{Angel}, we study the degradation of these proteins
using protocols 0 and I-C. We focus on the C-terminus because polyQ and
exon 1 of HTT are close to the N-terminus, so pulling from the N-terminus
is expected to be less favorable in comparison. The left panels of
Fig.~\ref{polyq_cv} show the probability of unfolding at force $F_{max}$ comparing
both protocols. Remarkably, the modal mechanical stability of Q$_{20}$ is significantly larger in an AFM-like
scenario (protocol 0, solid black) than when translocated through the proteasome (protocol I-C, dashed red).
On the other hand, the
distribution of mechanical stabilities for Q$_{60}$ in the proteasome
is shifted to the right and the number of non-mechanostable conformations ($F_{max}=0$) is significantly lower
than the case of the AFM-like pulling. This already suggests that pathological polyQ tracts are harder to
degrade than non-pathological ones.

S4~Fig. shows that there is no statistical relation
between the mechanical stability when measured by protocols 0 and I, as already
explored in ref.~\cite{Wojciechowski} for globular proteins. Furthermore,
knotted conformations are not clustered, but scattered and essentially
indistinguishable from the rest of the structures. Interestingly, two of the
knotted conformations show no force peaks in protocol 0 but have a high
force in the proteasome ($F_{max}\approx 1.8~\epsilon/$\AA{}), in
accordance with the results for globular knotted proteins.

We studied each of the Q$_{60}$ knotted conformations
from ref.~\cite{Angel} by translocating them through the proteasome at a
constant speed 20 times each. Most of the knotted conformations translocate
always, resulting only in a 7\% of jamming.

Upon the addition of the handles of HTT exon 1 to the knotted conformations,
nonetheless, there is a significant change in the behaviour: jamming increases
to an 11\%. The success rate in this case is then
reduced to 89\%. Examples of jamming and 
translocation can be observed as
force--extension plots in the middle panles of Fig.~\ref{polyq_cv}. The
right panels of this figure show differences between pulling simply the
polyQ region (Q$_{60}$) or the whole exon 1 of huntingtin (HTT$_{60}$). The mechanical
stabilities also show a significant shift: even though the fraction of
non-mechanostable conformations is similar, the distribution of forces for
HTT$_{60}$
is significantly shifted to the left,
which implies that smaller forces are needed in
order to unfold it than those needed for Q$_{60}$. Nonetheless, some 
structures show a large
increase in the force, reaching mechanical stabilities as high as
3.5~$\epsilon/$\AA{}. Interestingly, no correlation is observed between the
mechanical stability of HTT$_{60}$ and Q$_{60}$, suggesting that the handles 
may play a
very important role modulating this property.

\subsection*{Constant force}

The processes at constant $F_p$
are described by plots of the end-to-end distance, $L$, as a function of time, $t$, for
individual trajectories and by determining the associated characteristic success time.
For protocols 0 and II, we define success to be 
the unfolding of 85\% of the full backbone
length (when few, if any, contacts are left) and the characteristic time
is denoted by $t_U$.
For  protocol I, it is the translocation
time $t_{T}$. The native values of $L$ are 20.1, 22.0, 37.6, and 36.0~{\AA}
for 1J85, 1NS5, 1XD3, and 2EFV respectively.

\begin{figure}
\begin{center}
\includegraphics{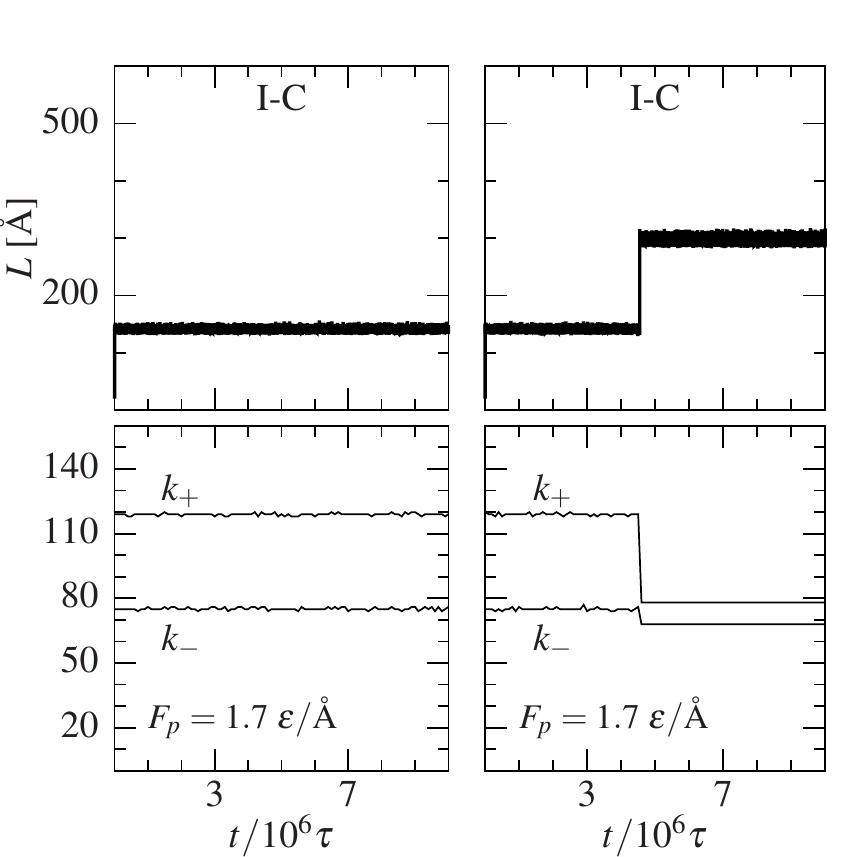}
\caption{\label{F1J85} 
Examples of constant force trajectories for 1J85 in the I-C protocol
for $F_p=1.7~\epsilon/${\AA}. Two situations arise for this protein:
The knot ends remain static (left) or the knot tightens (right), but translocation
is never achieved.
 }
\end{center}
\end{figure}

Fig.~\ref{F1J85} shows examples of the trajectories for 1J85 obtained under
protocol I-C. We observe no events of translocation, similar to the constant
velocity simulations (between 10 and 20 trajectories were generated for each
$F_p$ in this case). 
For $F_p \le 1.6~\epsilon/${\AA} and for most (90\%) 
of the trajectories for 1.7~$\epsilon/${\AA} the knot stays at its native location
as illustrated in the left panel of the figure. 
On the other hand, in all trajectories with
$F_p \ge 1.9~\epsilon/${\AA} 
and 45\% for 1.8~$\epsilon/${\AA} the knot gets
tightened -- both knot ends move to nearly the same locations.
The tightened knot may partially enter the channel of the 
proteasome model.
Similar results were obtained for 1NS5 in
protocols I-C and I-N, shown in
S5~Fig.: the knot may 
get tightened but there is no translocation.
We conclude that knots form an insurmountable  hindrance to
translocation of the deeply knotted proteins if pulling is
performed under the conditions of constant force.

Similarly, we have also studied Q$_{60}$ under a constant force unfolding
in the I-C configuration at different forces, and compared it to HTT$_{60}$. As
observed in constant velocity pulling, knotted conformations are sometimes
unable to translocate through the proteasome, leading to  stalling.
Right panels of Fig.~\ref{stuckvel} show snapshots of the translocation of one of the HTT
molecules, up to the untying of the knot.

\begin{figure}
\begin{center}
\includegraphics{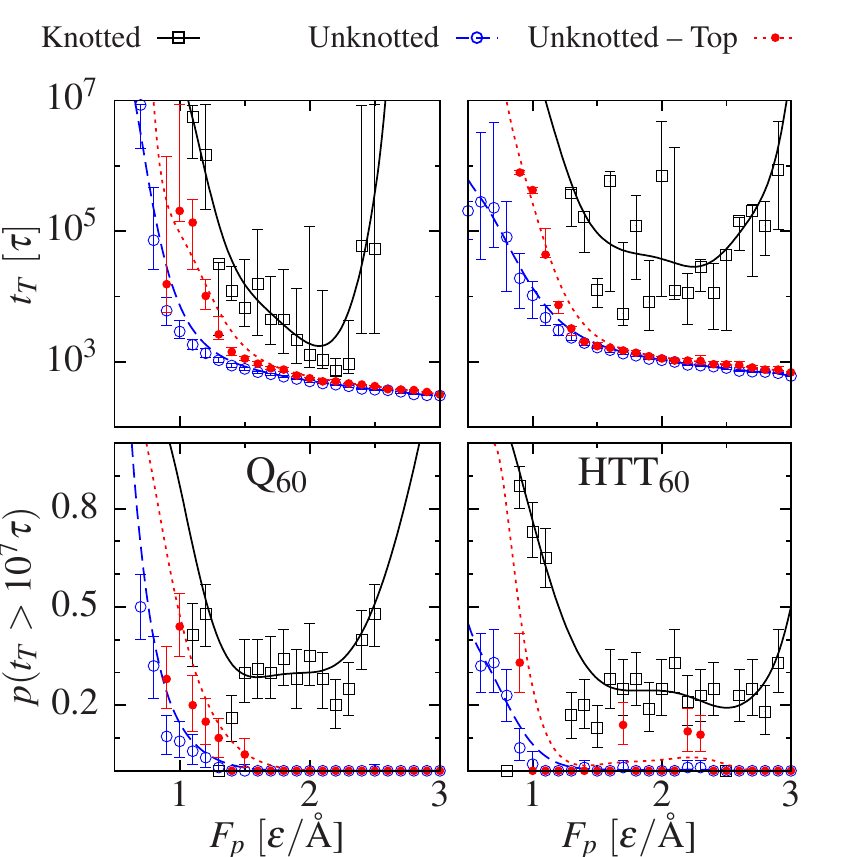}
\caption{\label{stalling} 
    Comparison of the translocation time (top) and stalling probability
    (bottom) of the knotted and unknotted Q$_{60}$ (left) and 
    HTT$_{60}$
    (right) as a function of $F_p$. The stalling probability is defined as the
    probability of the protein taking longer than $10^7~\tau$ to translocate.
    Each point corresponds to the median value, and the error bars represent a
    95\% confidence interval. It can be observed that knotted conformers (solid
    black line) take longer to translocate
    than unknotted ones (dashed blue line) and stall with a higher
    probability. Furthermore, even though mechanical stability does not relate
    to translocation time, the top-$F_{max}$ species (called
    ``Unknotted-Top'', dotted red line)
    translocate slower and stall more often than the average unknotted ones,
    although less often than knotted. The tendency at lower forces for both
    knotted and top-$F_{max}$ conformations is to stall the proteasome, so both
    could in principle be responsible for the malfunctioning of the proteasome.
    In this case, the knots are expected to be more troublesome, since they
    last for tenths of microseconds~\cite{Angel} as opposed to
nanoseconds~\cite{ns_fluctuations}.
 }
\end{center}
\end{figure}

The top panel of Fig.~\ref{stalling}
shows the translocation time as a function of the force for
Q$_{60}$ comparing unknotted conformations to knotted ones. 
The former show a
typical two-state curve for the force dependence whereby
$t_T$ depends on the
force exponentially in two regimes. The knotted conformations, however,
show an optimal $F_p$ for
translocation around 2~$\epsilon/$\AA{}. Forces below this value take longer to
untie the knot, while forces above  tend to stall the proteasome, so they take much
longer to translocate. 

The bottom panel of Fig.~\ref{stalling} shows the probability of the molecule
not being able to translocate in the simulation time ($10^7~\tau$), which
we operationally define as a stalling of the
proteasome. We can see that even if the translocation is not always
achieved at low forces, there is a significant difference between the stalling
of the proteasome due to knots and that due to other
mechanical elements such as the typical shearing mechanical clamps proposed
in~\cite{neurotos}.

Finally, 
the comparison of the isolated Q$_{60}$ to the one with exon 1 handles,
HTT$_{60}$, shows
that the translocation time grows, as expected by its
longer sequence, while the stalling is similar at high forces. Interestingly,
the N-terminal tail facilitates the unknotting at low forces, so that the
stalling is reduced -- compared to the isolated polyQ -- and is comparable to
the unknotted peptides. Furthermore, since it was
    proposed~\cite{neurotos} that proteasomal jamming might be induced by
    highly mechanostable conformations (called hM in ref.~\cite{neurotos}), we
    also compare the translocation time of the conformations that, in 
    protocol 0, have $F_{max}>1.5~\epsilon/$\AA{}. Indeed, the results shown
    in Fig.~\ref{stalling} show a significant difference between this
    subgroup and the whole set, but the effect due to knots is much greater.

Regarding the shorter polyQ chains, Q$_{40}$, our previous
study~\cite{Angel} did not generate conformations with knots --
most likely because of an insuffcient statistics.
Nonetheless, Q$_{40}$ is above the threshold of most polyglutaminopathies ($\sim$35) 
and above the minimal sequential extension of the trefoil knot.
Thus, the presence of knots in Q$_{40}$ should provide hindrance  to 
translocation like in the case of Q$_{60}$. In order to model this effect, 
we generated knotted Q$_{40}$ conformations by making use of the 
knotted conformations of length 60 and then by pruning the sequence at
both ends in such a way that the total length becomes 40 residues 
while the knot becomes centered (the number of the residues 
from $k_{-}$ to N is the same as the number of residues from
$k_{+}$ to C. S6~Fig. shows that the behavior of these
constructed knots is similar to the knotted subset of Q$_{60}$
at the relevant regime of smaller forces while is slightly different
at higher forces.

Taken together,  our results reinforce the hypothesis of the knots being responsible for
the malfunctioning of the degratation machinery, at least in polyQ-related
diseases, even if the degradation is stalled for shorter times.

\subsection*{Periodic force}

We now consider a situation in which the force is applied periodically:
for half of the period the force is $F_p$  while
for the other half the 
force is zero. We take the period to be 9000~$\tau$.

\begin{figure}
\begin{center}
\includegraphics{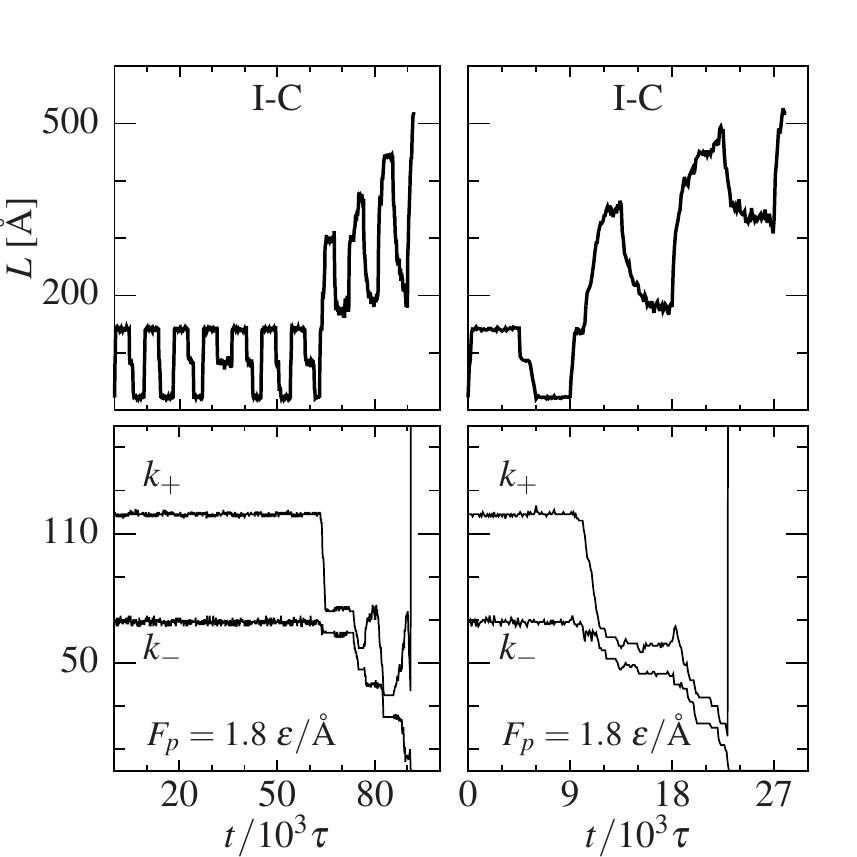}
\caption{\label{Periodic1NS5}
Examples of trajectories for 1NS5 with the pulling force applied periodically.
After each period of pulling, the protein may either retract (six
instances of
retraction to the near native situation in the left panels and one in the
right panels)
or stay in an extended state.
 }
\end{center}
\end{figure}

Fig.~\ref{Periodic1NS5} shows two examples of translocating trajectories 
for 1NS5 in protocol I-C at $F_p=1.8~\epsilon/${\AA}. 
In the initial stages of the process, $L$ gets extended but then
returns to the near-native value in the second idle part
of the periods. After a number of attempts, a substantially
longer extension arises and the return to the native situation
is no longer possible. Eventually, the translocation is accomplished.

\begin{figure}
    \centering
    \includegraphics{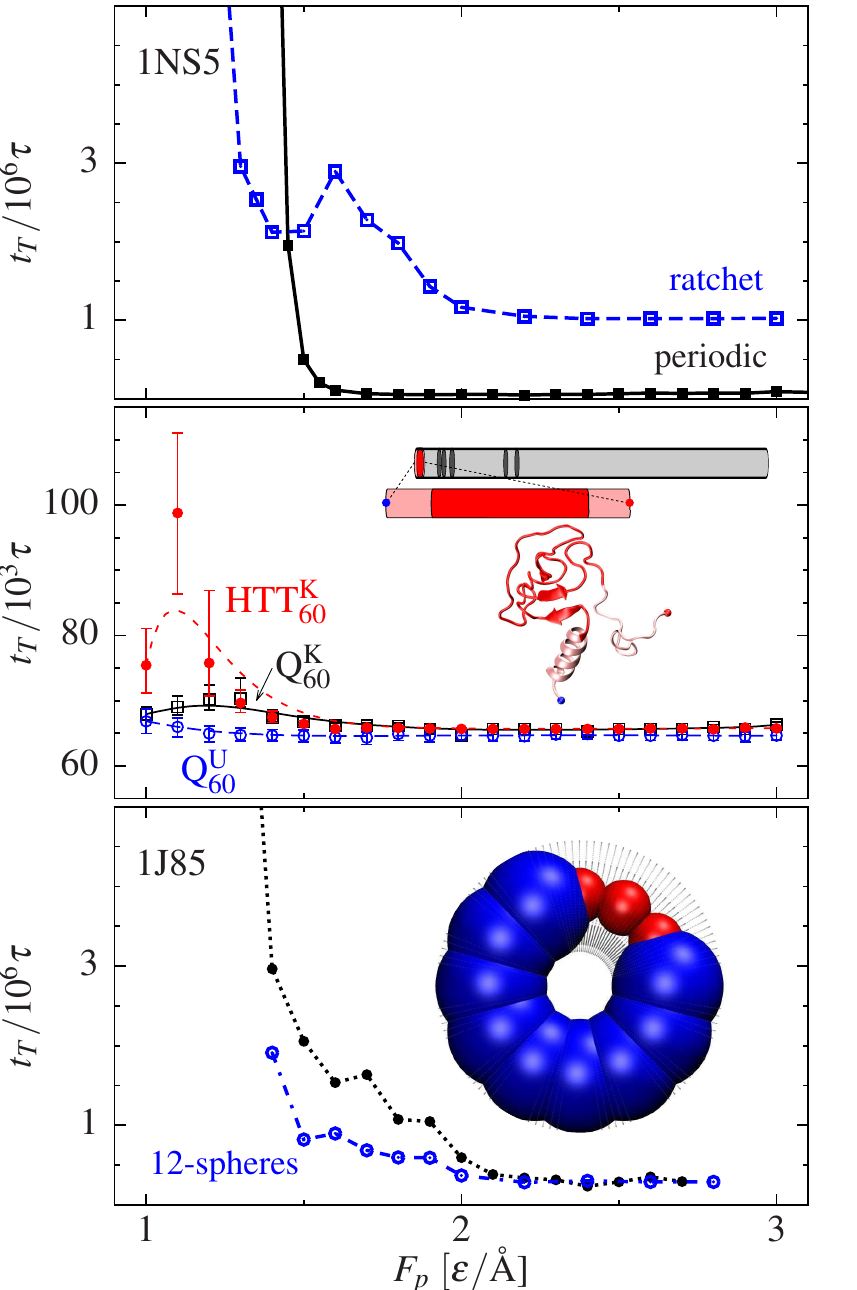}
    \caption{\label{periodic}
        Median translocation time as a function of $F_p$ for different models
        and systems. The top panel compares periodic pulling with retraction
        (black solid squares) to the ratchet model (which does not allow for
        retraction, blue open squares) on 1NS5.
        The middle panel compares the translocation of knotted and unknotted
        polyQ conformers (Q$^{\text{K}}_{60}$ in black and Q$^{\text{U}}_{60}$
        in blue, respectively) and knotted HTT$_{60}$ (HTT$^{\text{K}}_{60}$,
        in red) translocated using the ratchet model. The inset schematically
        shows the structure of Huntingtin and its exon 1, where the polyQ
        region is highlighted in red.
        The bottom panel shows the difference between the periodic model with a
        torus entrance (solid black circles) and one with an entrance of 12
        spheres of fixed size and a wider cylinder as channel (open blue
        circles), both of them with backtracking allowed. The inset
        schematically shows the 12-sphere model, in which three consecutive
        spheres shrink periodically (highlighted here in red).        
    }
\end{figure}

For a set of 100 trajectories, all are observed to lead to translocation at this $F_p$
and to involve untying of the knot. The median $t_{T}$ is 56~270~$\tau$.
On lowering the force, the median $t_{T}$ grows (the solid line in
the top panel of Fig.~\ref{periodic})
as does the number of instances in which the protein is stuck at the
entrance with its knot ends fixed at their native values (69 and 119).
For an $F_p$ of 1.40~$\epsilon/${\AA}, the odds of a successful translocation
become equal to jamming, at least within the cutoff time of $10^7~\tau$.
The shortest median $t_T$ is for an $F_p$ of 2.2~$\epsilon/${\AA}.
It increases to about 90~000~$\tau$ at 3.0~$\epsilon/${\AA} (which is hard
to see in the scale of the figure).

The periodic modulation of the pulling force imitates the 
periodic processing of the ATP molecules to the proteasome.
However, the exiting back out of the proteasome
does not seem to be observed experimentally~\cite{Bustamante}. In order
to remedy it, we consider a ratchet-like model in which the acceptable
advancement during one half-period is up to 3.8~{\AA}.
In the remaining part of the period, the protein in the channel is
held fixed by a harmonic spring at the pulling end
while the outside portion of the protein may equilibrate. 
The spring constant, $k_z$, used in the ratchet mechanism is weak and
is set to be equal to $2\; \epsilon$/{\AA}$^2$. In order to set the
reference location $z_{ref}$, we monitor the $z$-coordinate of the most 
forward residue. This provides the initial value of $z_{ref}$. If the
residue moves by 3.8~{\AA} with respect to $z_{ref}$ then the new
$z_{ref}$ is obtained by shifting the previous reference point by this
increment. The restoring force acting 
in the dwell phase, {\it i.e.}
when there is no dragging,
is equal to $k_z(z-z_{ref})$.

As a result, backtracking no longer takes place. 
As an unwanted byproduct, this mechanism
also prevents the rare events of refolding that have been observed
experimentally \cite{Sen,Maurizi}.
The dashed blue line in the top graph of Fig.~\ref{periodic} shows the median $t_T$ in the ratchet
model. The times are longer than 
in the absence of the ratchet in the regime of the
larger forces,
 which is also related to the fact that the distance pulled in one period
 is restricted. However, translocation still occurs for smaller forces, 
so it is expected that
in the biologically relevant regime of small forces,
of order 0.2~$\epsilon/${\AA}, the translocation becomes much more efficient.

In the periodic force pulling protocols, proteins may behave in a different
manner than in constant speed protocols. 
Considering, for example, proteins 1J85 and
1NS5, the former jams the proteasome during constant speed, but is degraded in
the periodic force protocol
while the latter jams the proteasome sometimes (68\%) at
constant speed, but is slightly worse processed in the periodic force protocol
than 1J85, see Fig.~\ref{periodic} -- bottom. It is interesting to note that, at least for
1J85, the knot-untying events take place usually during the
non-pulling part of the period when the protein attempts to refold.

The ratchet-like pulling model was also applied to
Q$_{60}$ and HTT$_{60}$, as shown in
the middle panel of Fig.~\ref{periodic}, which distinguishes between knotted and unknotted
conformations. In particular, the optimal pulling force of
2~$\varepsilon/$\AA{} determined for the constant force scenario is preserved
in the periodic pulling, even if the effect is much less dramatic in the
latter. Furthermore, we observe a maximum  similar to the case of 1NS5, but at a lower $F_p$
($\approx 1.2~\varepsilon/$\AA{}).
In any case, the translocation of knotted conformations
typically takes more time than the unknotted ones, and is expected to take
even longer at low forces, suggesting that the protein
degradation machinery might be stalled, or at least hindered, by the knotted conformations.

\subsection*{The 12-sphere model of the entrance to the proteasome}

The allosteric action of a proteasome involves not only traction
down the channel but also lateral fluctuational changes
in the shape of the intake opening~\cite{Tonddast}.
We now address the question of how to incorporate such rotation
in a model with an effective potential.

The front ring of the proteasome consists of six proteins
with loops that form the entrance to the funnel. These loops
undergo allosteric transformations resulting in a bending of the
individual loops and in a local
deformation of the shape of the entrance. 
These local deformations are not necessarily
consecutive around the entrance -- there is a strong random
component. However, there appears
to be a rotation-like correlation in the events~\cite{Tonddast}.
Our geometrical model may mimick this action by replacing
the torus by 12 circularly placed overlapping spheres of radius
6~{\AA} as illustrated in the inset in the bottom panel of Fig.~\ref{periodic}. The centers of the spheres
are located on a circle of radius $R_t$ (the major radius of the
original torus -- 13~{\AA}). The six loops are meant to be 
associated with every other sphere. The bending of a loop corresponds
to a temporal reduction in the radius of three  neighboring spheres
to 2~{\AA}, after which the radii return to their
larger values. For simplicity, we consider the
reductions to affect consecutive sets of three spheres
and the first sphere of the new set is taken to coincide
with the last sphere of the previous set, guaranteeing
the aforementioned association of a loop
with every other sphere. As a result, the
deformation in the shape of the entrance appears to be rotating.

This model requires to expand the cylindrical part of the funnel --
the radius of the cylinder is boosted from 8~to 12~{\AA} -- to allow 
for a smooth welding of the two parts of the funnel.
Otherwise, a gap between the cylinder and the smaller
sphere would form.
In the expanded channel, the protein may have enough space
to refold. In order to prevent this, we gradually reduce the
strength of the protein contacts in the channel to 0.
The molecular motor features of our model include the application
of a pulling force $F_p$ combined with the periodic
reduction of the radii of three spheres. We take a full
period (affecting all spheres consecutively) to be 6000~$\tau$ --
distinct from the 9000~$\tau$ periodicity of the axial force.

In S6~Fig. we show results for the unknotted
protein ADP-ribose pyrophosphatese (with the PDB code 2DSD; studied 
before in ref.~\cite{Wojciechowski}) translocating in the
model with the changing spheres. Compared to the model in which
the spheres do not change the radii, the translocation times
at low forces get shorter 
than in the torus-and-cylinder model 
(by about 84\% at an $F_p$ of 1.35~$\epsilon/${\AA}
and 20\% at 1.6~$\epsilon/${\AA}: the lower the force, the stronger the effect).

The reason for the larger effectiveness is that the unfolding
process at the entrance is helped by the time-varying shape of the 
funnel and translocation is improved by the wider cylinder.
We find (by freezing the spheres) that both effects are comparable
and become more important as $F_p$ is reduced. However, the enhancement by
moving spheres dominates at low forces.

Nonetheless, the rotational mechanism is found not to improve translocation
of deeply knotted proteins beyond what the model with non-changing 
spheres already does.
Unlike the smooth funnel model used at constant $v_p$ or $F_p$,
the model with the periodically applied force, with the spheres or
without, does allow for both jamming and translocating trajectories
for 1J85 in protocol I-C (see bottom panel of Fig.~\ref{periodic}).
In the periodic models with the fixed-size spheres (6~{\AA}),
the  median $t_T$ with the narrower cylinder is found to
be almost the same as in the smooth-funnel model.
On enlarging the cylinder, $t_T$ gets smaller, especially at 
lower forces. The interesting part is that adding the lateral
rotation related to the time-dependent radius of the spheres
does not bring any improvement.
The reason is that translocation of the knotted proteins does not
depend on any additional lateral forces but on the possibility
of untying in the idle part of each period.
The resulting misfolding during the idle times
leads occasionaly to the pulled segment  to return without threading
through the knot-loop, {\it i.e.} to untying. 

The 12-sphere mod\-el, es\-pe\-cial\-ly
with the ratch\-et-like block\-ing mech\-an\-ism, is probably the most realistic
of the set of models considered here. However, its proper working depends on 
the adequate choice of the parameters. 
For instance, when we implement the ratchet mechanism,
we impose a condition on the maximum translocation allowed in one
period. We take it as 3.8~{\AA}, which probably is non-optimal.
A greater length 
may lead to a faster translocation, especially of 1J85
-- a process which is not very efficient at low forces. 
More generally, this length may depend on $F_p$.

\subsection*{Constant force -- all-atom simulations}

\begin{figure}
\begin{center}
\includegraphics{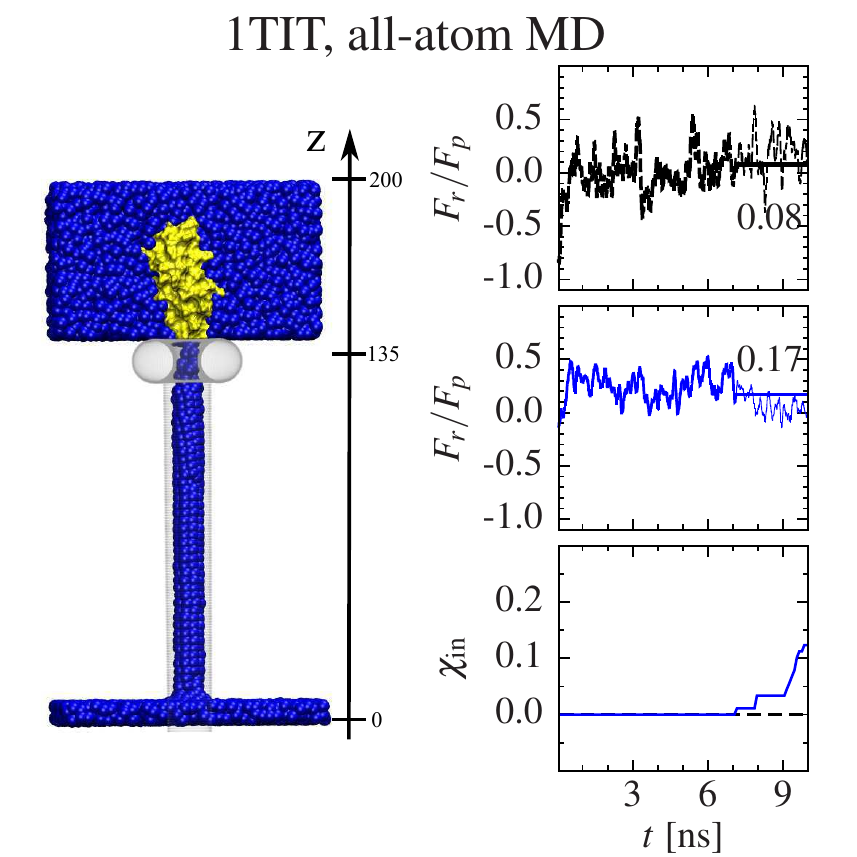}
\caption{\label{allatomend} 
Results of the all-atom simulations for 1TIT. The left section shows the model
used for the simulation. At the right, the top and middle panels show the 
ratio of measured force to applied force when the latter is 2.3 and 
4.5~$\epsilon/$\AA{}, respectively. The bottom panel shows the fraction of 
residues that enter the channel as a function of time.
 }
\end{center}
\end{figure}

Fig.~\ref{allatomend} shows the results pertaining to pulling 1TIT under 
protocol II-C  using all-atom simulations.
In order to neutralize the net charge of -6~$e$ 
and to have the ionic strength
of about 150~mM, we add 64~Na$^+$ and 58~Cl$^-$ ions.
Without the ions, translocation is somewhat faster.

We have focused on two values of $F_p$: 2.3 and 4.5~$\epsilon/${\AA}.
For larger $F_p$ (such as 15~$\epsilon/${\AA})
translocation  is instantaneous and
for $F_p< 2.3~\epsilon/${\AA} there is no forward motion within the
time scale of the simulations.
After the initial transients, the ratio of the tension at the back terminus
to the pulling force is observed to be 0.08 and 0.17 for $F_p$ of
2.3 and 4.5~$\epsilon/${\AA} respectively,
confirming the mechanical role of the proteasome in defining the
conditions of equilibrium. 
Similar results are obtained for several rotated
orientations of the protein about the main axis.
At the larger force, about 10\% of the
residues are seen observed to translocate through the reference plane.
Removal of all ions does not affect
the tension at the back terminus.

\section*{Conclusions}
In a previous study~\cite{Wojciechowski} we have found that the ease of
unfolding and translocation depends on the protein, on the protocol of  
pulling and on the value of $F_p$. The same is true in the case of proteins
with knots, but the added complexity of such proteins 
results in either jamming of
the proteasome or, at least, an extension of
the translocation times significantly.
The hindrances are found to grow more and more 
powerful on lowering
$F_p$ to the typical values of the
biological motors.
The shallowly knotted proteins and the knotted polyQ structures are more
likely to hinder the translocation process than to block it,
but the hindrances grow rapidly on lowering the pulling force 
to the biologically relevant regime.
On the other hand, the deeply knotted proteins are more likely to jam,
especially at small forces (smaller than those used in our simulations)
which are relevant biologically. These results, qualitatively, do not depend 
on the version of the model of the proteasome that is used,
even though the timescales do depend on the model. However, the specific outcome
depends on the particular model used, for instance, on whether the
force is applied periodically or if the ratchet-like blocking mechanism
is built in. 

Recently, Szymczak~\cite{Szymczaktrans,Szymczakspecial} has considered
translocation of several knotted proteins through a cylindrical pore connected
to a flat plane which is meant to relate to transport through cellular
membranes such as the ones in mitochodria. 
In his model, the interactions with the pore wall
are different than in our model 
but this author 
has also observed a 
variety of protein-dependent behaviors when a pulling force was
applied
and noted that jamming could be avoided by
making the force act periodically~\cite{Szymczakperiodic}.

On the experimental side, Jackson {\it et al.}
have considered knotted proteins 1XD3 and 1NS5 in ClpP/X assays and
also found a rich variety in their behavior 
(personal communication).
1NS5 entering through the C-terminus
degrades rapidly, consistent with our results with the periodic
force, unless a stable ThiS domain is attached at the N-terminus.
On the other hand, 1XD3 is resistant to degradation in what we call
protocol I-C.  This is surprising because the N-terminal knot is
shallow and should untie readily as we find in the
constant-speed case.

Specifically in the case of neurodegeneration, our work shows how knotted
polyglutamine tracts hinder the protesomal function both when isolated and when
flanked by one of the naturally occurring flanking sequences, huntingtin exon
1. An inefficient degradation may unbalance the concentration of elements
to degrade in the cell, which might be related
to toxcicity in two
ways: the acumulation of monomers will increase the probability of one of
them becoming toxic; and the enhanced number of molecules will increase
the concentration of the neurotoxic
proteins, which will then aggregate into toxic oligomers after the critical
concentration of the process is reached.  Both mechanisms of toxicity
suggested for polyglutamine differ from the proposed effect of metals, such as
copper and zinc~\cite{Sensi,Kozlowski,Ruth,Viles} in the aggregation of
\ga{}-synuclein or \gb{}-amyloid. This marks an important difference
between the genetically determined diseases (polyglutaminopathies such as
Huntington) compared to environmentally driven ones (like the sporadic form of
Alzheimer or Parkinson).

There has been considerable debate about the role of knots in proteins in general.
One possible role is to enhance the mechanical, kinetic, and thermodynamic
stability of a protein~\cite{stability}. Here, we have investigated a harmful
role: reduction in the efficiency or even derailing of the protein degradation 
process, which may result in toxicity.

\section*{Acknowledgments}
We appreciate stimulating discussions with S. E. Jackson, M. Sikora, R.
Herv\'as and
K. Wo\l{}ek. We also acknowledge the contribution of SVG High-Performance Computing facility
provided by the Galician Supercomputing Center (CESGA), \texttt{www.cesga.es}.
 This work has been supported by the EU Joint Programme in
Neurodegenerative Diseases project (JPND CD FP-688-059). The project is
supported through the following funding organisations under the aegis of
JPND - \texttt{www.jpnd.eu}: Ireland, HRB; Poland, National Science 
Centre (2014/15/Z/NZ1/00037); and
Spain, ISCIII (AC14/00037 ISCIII). MW has been supported by Polish National Science Centre 
Grant No. 2014/15/B/ST3/01905. AGS was supported by a JAE
doctoral fellowship from CSIC.






\begin{mcitethebibliography}{75}
\providecommand*{\natexlab}[1]{#1}
\providecommand*{\mciteSetBstSublistMode}[1]{}
\providecommand*{\mciteSetBstMaxWidthForm}[2]{}
\providecommand*{\mciteBstWouldAddEndPuncttrue}
  {\def\EndOfBibitem{\unskip.}}
\providecommand*{\mciteBstWouldAddEndPunctfalse}
  {\let\EndOfBibitem\relax}
\providecommand*{\mciteSetBstMidEndSepPunct}[3]{}
\providecommand*{\mciteSetBstSublistLabelBeginEnd}[3]{}
\providecommand*{\EndOfBibitem}{}
\mciteSetBstSublistMode{f}
\mciteSetBstMaxWidthForm{subitem}
{(\emph{\alph{mcitesubitemcount}})}
\mciteSetBstSublistLabelBeginEnd{\mcitemaxwidthsubitemform\space}
{\relax}{\relax}

\bibitem[Mulligan and Chakrabartty(2013)]{diseases_of_the_old}
V.~K. Mulligan and A.~Chakrabartty, \emph{Proteins Struct Funct Bioinf}, 2013,
  \textbf{81}, 1285--1303\relax
\mciteBstWouldAddEndPuncttrue
\mciteSetBstMidEndSepPunct{\mcitedefaultmidpunct}
{\mcitedefaultendpunct}{\mcitedefaultseppunct}\relax
\EndOfBibitem
\bibitem[Sawaya \emph{et~al.}(2007)Sawaya, Sambashivan, Nelson, Ivanova,
  Sievers, Apostol, Thompson, Balbirnie, Wiltzius,
  McFarlane,\emph{et~al.}]{cross_beta_spine}
M.~R. Sawaya, S.~Sambashivan, R.~Nelson, M.~I. Ivanova, S.~A. Sievers, M.~I.
  Apostol, M.~J. Thompson, M.~Balbirnie, J.~J. Wiltzius, H.~T. McFarlane
  \emph{et~al.}, \emph{Nature}, 2007, \textbf{447}, 453--457\relax
\mciteBstWouldAddEndPuncttrue
\mciteSetBstMidEndSepPunct{\mcitedefaultmidpunct}
{\mcitedefaultendpunct}{\mcitedefaultseppunct}\relax
\EndOfBibitem
\bibitem[Kayed \emph{et~al.}(2003)Kayed, Head, Thompson, McIntire, Milton,
  Cotman, and Glabe]{a11}
R.~Kayed, E.~Head, J.~L. Thompson, T.~M. McIntire, S.~C. Milton, C.~W. Cotman
  and C.~G. Glabe, \emph{Science}, 2003, \textbf{300}, 486--489\relax
\mciteBstWouldAddEndPuncttrue
\mciteSetBstMidEndSepPunct{\mcitedefaultmidpunct}
{\mcitedefaultendpunct}{\mcitedefaultseppunct}\relax
\EndOfBibitem
\bibitem[Herv\'as \emph{et~al.}(2012)Herv\'as, Oroz, Galera-Prat, {Go\~{n}i},
  Valbuena, Vera, G\'omez-Sicilia, Losada-Urz\'aiz, Uversky, Men\'endez,
  Laurents, Bruix, and Carri\'on-V\'azquez]{neurotos}
R.~Herv\'as, J.~Oroz, A.~Galera-Prat, O.~{Go\~{n}i}, A.~Valbuena, A.~M. Vera,
  A.~G\'omez-Sicilia, F.~Losada-Urz\'aiz, V.~N. Uversky, M.~Men\'endez, D.~V.
  Laurents, M.~Bruix and M.~Carri\'on-V\'azquez, \emph{PLoS Biol}, 2012,
  \textbf{10}, e1001335\relax
\mciteBstWouldAddEndPuncttrue
\mciteSetBstMidEndSepPunct{\mcitedefaultmidpunct}
{\mcitedefaultendpunct}{\mcitedefaultseppunct}\relax
\EndOfBibitem
\bibitem[Ferreon \emph{et~al.}(2010)Ferreon, Moran, Gambin, and
  Deniz]{ns_fluctuations}
A.~C.~M. Ferreon, C.~R. Moran, Y.~Gambin and A.~A. Deniz, \emph{Methods
  Enzymol}, 2010, \textbf{472}, 179--204\relax
\mciteBstWouldAddEndPuncttrue
\mciteSetBstMidEndSepPunct{\mcitedefaultmidpunct}
{\mcitedefaultendpunct}{\mcitedefaultseppunct}\relax
\EndOfBibitem
\bibitem[G{\'{o}}mez-Sicilia \emph{et~al.}(2015)G{\'{o}}mez-Sicilia, Sikora,
  Cieplak, and Carri{\'{o}}n-V{\'{a}}zquez]{Angel}
{\`{A}}.~G{\'{o}}mez-Sicilia, M.~Sikora, M.~Cieplak and
  M.~Carri{\'{o}}n-V{\'{a}}zquez, \emph{PLoS Comput Biol}, 2015, \textbf{11},
  e1004541\relax
\mciteBstWouldAddEndPuncttrue
\mciteSetBstMidEndSepPunct{\mcitedefaultmidpunct}
{\mcitedefaultendpunct}{\mcitedefaultseppunct}\relax
\EndOfBibitem
\bibitem[Cossio \emph{et~al.}(2010)Cossio, Trovato, Pietrucci, Seno, Maritan,
  and Laio]{val60}
P.~Cossio, A.~Trovato, F.~Pietrucci, F.~Seno, A.~Maritan and A.~Laio,
  \emph{PLoS Comput Biol}, 2010, \textbf{6}, e1000957\relax
\mciteBstWouldAddEndPuncttrue
\mciteSetBstMidEndSepPunct{\mcitedefaultmidpunct}
{\mcitedefaultendpunct}{\mcitedefaultseppunct}\relax
\EndOfBibitem
\bibitem[Sillitoe \emph{et~al.}(2013)Sillitoe, Cuff, Dessailly, Dawson,
  Furnham, Lee, Lees, Lewis, Studer, Rentzsch, Yeats, Thornton, and
  Orengo]{cath}
I.~Sillitoe, A.~L. Cuff, B.~H. Dessailly, N.~L. Dawson, N.~Furnham, D.~Lee,
  J.~G. Lees, T.~E. Lewis, R.~A. Studer, R.~Rentzsch, C.~Yeats, J.~M. Thornton
  and C.~A. Orengo, \emph{Nucleic Acids Res}, 2013, \textbf{41},
  D490--D498\relax
\mciteBstWouldAddEndPuncttrue
\mciteSetBstMidEndSepPunct{\mcitedefaultmidpunct}
{\mcitedefaultendpunct}{\mcitedefaultseppunct}\relax
\EndOfBibitem
\bibitem[Jamroz \emph{et~al.}(2015)Jamroz, Niemyska, Rawdon, Stasiak, Millett,
  Su{\l}kowski, and Sulkowska]{Sulkowskabaza}
M.~Jamroz, W.~Niemyska, E.~J. Rawdon, A.~Stasiak, K.~C. Millett,
  P.~Su{\l}kowski and J.~I. Sulkowska, \emph{Nucleic Acids Res}, 2015,
  \textbf{43}, D306--D314\relax
\mciteBstWouldAddEndPuncttrue
\mciteSetBstMidEndSepPunct{\mcitedefaultmidpunct}
{\mcitedefaultendpunct}{\mcitedefaultseppunct}\relax
\EndOfBibitem
\bibitem[Virnau \emph{et~al.}(2006)Virnau, Mirny, and Kardar]{knots2}
P.~Virnau, L.~A. Mirny and M.~Kardar, \emph{PLoS Comput Biol}, 2006,
  \textbf{2}, e122\relax
\mciteBstWouldAddEndPuncttrue
\mciteSetBstMidEndSepPunct{\mcitedefaultmidpunct}
{\mcitedefaultendpunct}{\mcitedefaultseppunct}\relax
\EndOfBibitem
\bibitem[a~Beccara \emph{et~al.}(2013)a~Beccara, \v{S}krbi\'c, Covino,
  Micheletti, and Faccioli]{Micheletti}
S.~a~Beccara, T.~\v{S}krbi\'c, R.~Covino, C.~Micheletti and P.~Faccioli,
  \emph{PLoS Comput Biol}, 2013, \textbf{9}, e1003002\relax
\mciteBstWouldAddEndPuncttrue
\mciteSetBstMidEndSepPunct{\mcitedefaultmidpunct}
{\mcitedefaultendpunct}{\mcitedefaultseppunct}\relax
\EndOfBibitem
\bibitem[Noel \emph{et~al.}(2013)Noel, Onuchic, and Sulkowska]{shallowJoanna}
J.~K. Noel, J.~N. Onuchic and J.~I. Sulkowska, \emph{J Phys Chem Lett}, 2013,
  \textbf{4}, 3570--3573\relax
\mciteBstWouldAddEndPuncttrue
\mciteSetBstMidEndSepPunct{\mcitedefaultmidpunct}
{\mcitedefaultendpunct}{\mcitedefaultseppunct}\relax
\EndOfBibitem
\bibitem[Chwastyk and Cieplak(2015)]{knotshallow}
M.~Chwastyk and M.~Cieplak, \emph{J Chem Phys}, 2015, \textbf{143},
  045101\relax
\mciteBstWouldAddEndPuncttrue
\mciteSetBstMidEndSepPunct{\mcitedefaultmidpunct}
{\mcitedefaultendpunct}{\mcitedefaultseppunct}\relax
\EndOfBibitem
\bibitem[Berman \emph{et~al.}(2000)Berman, Westbrook, Feng, Gilliland, Bhat,
  Weissig, Shindyalov, and Bourne]{pdb}
H.~M. Berman, J.~Westbrook, Z.~Feng, G.~Gilliland, T.~N. Bhat, H.~Weissig,
  I.~N. Shindyalov and P.~E. Bourne, \emph{Nucleic Acids Res}, 2000,
  \textbf{28}, 235--242\relax
\mciteBstWouldAddEndPuncttrue
\mciteSetBstMidEndSepPunct{\mcitedefaultmidpunct}
{\mcitedefaultendpunct}{\mcitedefaultseppunct}\relax
\EndOfBibitem
\bibitem[Virnau \emph{et~al.}(2011)Virnau, Mallam, and Jackson]{VirnauJackson}
P.~Virnau, A.~Mallam and S.~Jackson, \emph{J Phys Condens Matter}, 2011,
  \textbf{23}, 033101\relax
\mciteBstWouldAddEndPuncttrue
\mciteSetBstMidEndSepPunct{\mcitedefaultmidpunct}
{\mcitedefaultendpunct}{\mcitedefaultseppunct}\relax
\EndOfBibitem
\bibitem[Su{\l}kowska \emph{et~al.}(2012)Su{\l}kowska, Rawdon, Millett,
  Onuchic, and Stasiak]{Sulkowska}
J.~I. Su{\l}kowska, E.~J. Rawdon, K.~C. Millett, J.~N. Onuchic and A.~Stasiak,
  \emph{Proc Natl Acad Sci USA}, 2012, \textbf{109}, E1715--E1723\relax
\mciteBstWouldAddEndPuncttrue
\mciteSetBstMidEndSepPunct{\mcitedefaultmidpunct}
{\mcitedefaultendpunct}{\mcitedefaultseppunct}\relax
\EndOfBibitem
\bibitem[Lim and Jackson(2015)]{LimJackson}
N.~C.~H. Lim and S.~E. Jackson, \emph{J Phys Condens Matter}, 2015,
  \textbf{27}, 354101\relax
\mciteBstWouldAddEndPuncttrue
\mciteSetBstMidEndSepPunct{\mcitedefaultmidpunct}
{\mcitedefaultendpunct}{\mcitedefaultseppunct}\relax
\EndOfBibitem
\bibitem[Konrat(2015)]{Konrat}
R.~Konrat, \emph{Biophys J}, 2015, \textbf{109}, 1309--1311\relax
\mciteBstWouldAddEndPuncttrue
\mciteSetBstMidEndSepPunct{\mcitedefaultmidpunct}
{\mcitedefaultendpunct}{\mcitedefaultseppunct}\relax
\EndOfBibitem
\bibitem[Nasir \emph{et~al.}(1995)Nasir, Floresco, O'Kusky, Diewert, Richman,
  Zeisler, Borowski, Marth, Phillips, and Hayden]{htt_development}
J.~Nasir, S.~B. Floresco, J.~R. O'Kusky, V.~M. Diewert, J.~M. Richman,
  J.~Zeisler, A.~Borowski, J.~D. Marth, A.~G. Phillips and M.~R. Hayden,
  \emph{Cell}, 1995, \textbf{81}, 811 -- 823\relax
\mciteBstWouldAddEndPuncttrue
\mciteSetBstMidEndSepPunct{\mcitedefaultmidpunct}
{\mcitedefaultendpunct}{\mcitedefaultseppunct}\relax
\EndOfBibitem
\bibitem[Zuccato \emph{et~al.}(2001)Zuccato, Ciammola, Rigamonti, Leavitt,
  Goffredo, Conti, MacDonald, Friedlander, Silani, Hayden, Timmusk, Sipione,
  and Cattaneo]{htt_gene_expression}
C.~Zuccato, A.~Ciammola, D.~Rigamonti, B.~R. Leavitt, D.~Goffredo, L.~Conti,
  M.~E. MacDonald, R.~M. Friedlander, V.~Silani, M.~R. Hayden, T.~Timmusk,
  S.~Sipione and E.~Cattaneo, \emph{Science}, 2001, \textbf{293},
  493--498\relax
\mciteBstWouldAddEndPuncttrue
\mciteSetBstMidEndSepPunct{\mcitedefaultmidpunct}
{\mcitedefaultendpunct}{\mcitedefaultseppunct}\relax
\EndOfBibitem
\bibitem[Velier \emph{et~al.}(1998)Velier, Kim, Schwarz, Kim, Sapp, Chase,
  Aronin, and DiFiglia]{htt_vessicles}
J.~Velier, M.~Kim, C.~Schwarz, T.~W. Kim, E.~Sapp, K.~Chase, N.~Aronin and
  M.~DiFiglia, \emph{Exp Neurology}, 1998, \textbf{152}, 34 -- 40\relax
\mciteBstWouldAddEndPuncttrue
\mciteSetBstMidEndSepPunct{\mcitedefaultmidpunct}
{\mcitedefaultendpunct}{\mcitedefaultseppunct}\relax
\EndOfBibitem
\bibitem[Petruska \emph{et~al.}(1998)Petruska, Hartenstine, and
  Goodman]{polymerase_slippage}
J.~Petruska, M.~J. Hartenstine and M.~F. Goodman, \emph{J Biol Chem}, 1998,
  \textbf{273}, 5204--5210\relax
\mciteBstWouldAddEndPuncttrue
\mciteSetBstMidEndSepPunct{\mcitedefaultmidpunct}
{\mcitedefaultendpunct}{\mcitedefaultseppunct}\relax
\EndOfBibitem
\bibitem[Warby \emph{et~al.}(2011)Warby, Visscher, Collins, Doty, Carter,
  Butland, Hayden, Kanazawa, Ross, and Hayden]{Warby}
S.~C. Warby, H.~Visscher, J.~A. Collins, C.~N. Doty, C.~Carter, S.~L. Butland,
  A.~R. Hayden, I.~Kanazawa, C.~J. Ross and M.~R. Hayden, \emph{Eur J Hum
  Genet}, 2011, \textbf{19}, 561--566\relax
\mciteBstWouldAddEndPuncttrue
\mciteSetBstMidEndSepPunct{\mcitedefaultmidpunct}
{\mcitedefaultendpunct}{\mcitedefaultseppunct}\relax
\EndOfBibitem
\bibitem[DiFiglia \emph{et~al.}(1997)DiFiglia, Sapp, Chase, Davies, Bates,
  Vonsattel, and Aronin]{DiFiglia}
M.~DiFiglia, E.~Sapp, K.~O. Chase, S.~W. Davies, G.~P. Bates, J.~P. Vonsattel
  and N.~Aronin, \emph{Science}, 1997, \textbf{277}, 1990--1993\relax
\mciteBstWouldAddEndPuncttrue
\mciteSetBstMidEndSepPunct{\mcitedefaultmidpunct}
{\mcitedefaultendpunct}{\mcitedefaultseppunct}\relax
\EndOfBibitem
\bibitem[Verma \emph{et~al.}(2015)Verma, Vats, and Taneja]{toxic_oligomers1}
M.~Verma, A.~Vats and V.~Taneja, \emph{Ann Indian Acad Neurol}, 2015,
  \textbf{18}, 138--145\relax
\mciteBstWouldAddEndPuncttrue
\mciteSetBstMidEndSepPunct{\mcitedefaultmidpunct}
{\mcitedefaultendpunct}{\mcitedefaultseppunct}\relax
\EndOfBibitem
\bibitem[Ross and Poirier(2004)]{toxic_oligomers2}
C.~A. Ross and M.~A. Poirier, \emph{Nat Med}, 2004, \textbf{10 Suppl},
  S10--S17\relax
\mciteBstWouldAddEndPuncttrue
\mciteSetBstMidEndSepPunct{\mcitedefaultmidpunct}
{\mcitedefaultendpunct}{\mcitedefaultseppunct}\relax
\EndOfBibitem
\bibitem[Nagai \emph{et~al.}(2007)Nagai, Inui, Popiel, Fujikake, Hasegawa,
  Urade, Goto, Naiki, and Toda]{nagai_toxic_monomer}
Y.~Nagai, T.~Inui, H.~A. Popiel, N.~Fujikake, K.~Hasegawa, Y.~Urade, Y.~Goto,
  H.~Naiki and T.~Toda, \emph{Nat Struct Mol Biol}, 2007, \textbf{14},
  332--340\relax
\mciteBstWouldAddEndPuncttrue
\mciteSetBstMidEndSepPunct{\mcitedefaultmidpunct}
{\mcitedefaultendpunct}{\mcitedefaultseppunct}\relax
\EndOfBibitem
\bibitem[Bence \emph{et~al.}(2001)Bence, Sampat, and
  Kopito]{impairment_proteasome}
N.~F. Bence, R.~M. Sampat and R.~R. Kopito, \emph{Science}, 2001, \textbf{292},
  1552--1555\relax
\mciteBstWouldAddEndPuncttrue
\mciteSetBstMidEndSepPunct{\mcitedefaultmidpunct}
{\mcitedefaultendpunct}{\mcitedefaultseppunct}\relax
\EndOfBibitem
\bibitem[Koniaris and Muthukumar(1991)]{km1}
Koniaris and Muthukumar, \emph{Phys Rev Lett}, 1991, \textbf{66},
  2211--2214\relax
\mciteBstWouldAddEndPuncttrue
\mciteSetBstMidEndSepPunct{\mcitedefaultmidpunct}
{\mcitedefaultendpunct}{\mcitedefaultseppunct}\relax
\EndOfBibitem
\bibitem[Taylor(2000)]{taylor}
W.~R. Taylor, \emph{Nature}, 2000, \textbf{406}, 916--919\relax
\mciteBstWouldAddEndPuncttrue
\mciteSetBstMidEndSepPunct{\mcitedefaultmidpunct}
{\mcitedefaultendpunct}{\mcitedefaultseppunct}\relax
\EndOfBibitem
\bibitem[Chwastyk and Cieplak(2015)]{chwastdeep}
M.~Chwastyk and M.~Cieplak, \emph{J Phys Condens Matter}, 2015, \textbf{27},
  354105\relax
\mciteBstWouldAddEndPuncttrue
\mciteSetBstMidEndSepPunct{\mcitedefaultmidpunct}
{\mcitedefaultendpunct}{\mcitedefaultseppunct}\relax
\EndOfBibitem
\bibitem[Wojciechowski \emph{et~al.}(2014)Wojciechowski, Szymczak,
  Carri{\'{o}}n-V{\'{a}}zquez, and Cieplak]{Wojciechowski}
M.~Wojciechowski, P.~Szymczak, M.~Carri{\'{o}}n-V{\'{a}}zquez and M.~Cieplak,
  \emph{Biophys J}, 2014, \textbf{107}, 1661--1668\relax
\mciteBstWouldAddEndPuncttrue
\mciteSetBstMidEndSepPunct{\mcitedefaultmidpunct}
{\mcitedefaultendpunct}{\mcitedefaultseppunct}\relax
\EndOfBibitem
\bibitem[Coux \emph{et~al.}(1996)Coux, Tanaka, and Goldberg]{Coux_1996}
O.~Coux, K.~Tanaka and A.~L. Goldberg, \emph{Annu Rev Biochem}, 1996,
  \textbf{65}, 801--847\relax
\mciteBstWouldAddEndPuncttrue
\mciteSetBstMidEndSepPunct{\mcitedefaultmidpunct}
{\mcitedefaultendpunct}{\mcitedefaultseppunct}\relax
\EndOfBibitem
\bibitem[Goldberg(1990)]{Goldberg_1990}
A.~L. Goldberg, \emph{Semin Cell Biol}, 1990, \textbf{1}, 423--432\relax
\mciteBstWouldAddEndPuncttrue
\mciteSetBstMidEndSepPunct{\mcitedefaultmidpunct}
{\mcitedefaultendpunct}{\mcitedefaultseppunct}\relax
\EndOfBibitem
\bibitem[Gottesman(1996)]{Gottesman_1996}
S.~Gottesman, \emph{Annu Rev Genet}, 1996, \textbf{30}, 465--506\relax
\mciteBstWouldAddEndPuncttrue
\mciteSetBstMidEndSepPunct{\mcitedefaultmidpunct}
{\mcitedefaultendpunct}{\mcitedefaultseppunct}\relax
\EndOfBibitem
\bibitem[Zhang \emph{et~al.}(2009)Zhang, Hu, Tian, Zhang, Finley, Jeffrey, and
  Shi]{Zhang_2009_struct}
F.~Zhang, M.~Hu, G.~Tian, P.~Zhang, D.~Finley, P.~D. Jeffrey and Y.~Shi,
  \emph{Mol Cell}, 2009, \textbf{34}, 473 -- 484\relax
\mciteBstWouldAddEndPuncttrue
\mciteSetBstMidEndSepPunct{\mcitedefaultmidpunct}
{\mcitedefaultendpunct}{\mcitedefaultseppunct}\relax
\EndOfBibitem
\bibitem[Groll \emph{et~al.}(1997)Groll, Ditzel, L{\"{o}}we, Stock, Bochtler,
  Bartunik, and Huber]{Groll_1997}
M.~Groll, L.~Ditzel, J.~L{\"{o}}we, D.~Stock, M.~Bochtler, H.~D. Bartunik and
  R.~Huber, \emph{Nature}, 1997, \textbf{386}, 463--471\relax
\mciteBstWouldAddEndPuncttrue
\mciteSetBstMidEndSepPunct{\mcitedefaultmidpunct}
{\mcitedefaultendpunct}{\mcitedefaultseppunct}\relax
\EndOfBibitem
\bibitem[Kravats \emph{et~al.}(2011)Kravats, Jayasinghe, and Stan]{Stan}
A.~Kravats, M.~Jayasinghe and G.~Stan, \emph{Proc Natl Acad Sci USA}, 2011,
  \textbf{108}, 2234--2239\relax
\mciteBstWouldAddEndPuncttrue
\mciteSetBstMidEndSepPunct{\mcitedefaultmidpunct}
{\mcitedefaultendpunct}{\mcitedefaultseppunct}\relax
\EndOfBibitem
\bibitem[Tonddast-Navaei and Stan(2013)]{Tonddast}
S.~Tonddast-Navaei and G.~Stan, \emph{J Am Chem Soc}, 2013, \textbf{135},
  14627--14636\relax
\mciteBstWouldAddEndPuncttrue
\mciteSetBstMidEndSepPunct{\mcitedefaultmidpunct}
{\mcitedefaultendpunct}{\mcitedefaultseppunct}\relax
\EndOfBibitem
\bibitem[Kravats \emph{et~al.}(2013)Kravats, Tonddast-Navaei, Bucher, and
  Stan]{Kravats-Tonddast}
A.~N. Kravats, S.~Tonddast-Navaei, R.~J. Bucher and G.~Stan, \emph{J Chem
  Phys}, 2013, \textbf{139}, 121921\relax
\mciteBstWouldAddEndPuncttrue
\mciteSetBstMidEndSepPunct{\mcitedefaultmidpunct}
{\mcitedefaultendpunct}{\mcitedefaultseppunct}\relax
\EndOfBibitem
\bibitem[Maillard \emph{et~al.}(2011)Maillard, Chistol, Sen, Righini, Tan,
  Kaiser, Hodges, Martin, and Bustamante]{Bustamante}
R.~A. Maillard, G.~Chistol, M.~Sen, M.~Righini, J.~Tan, C.~M. Kaiser,
  C.~Hodges, A.~Martin and C.~Bustamante, \emph{Cell}, 2011, \textbf{145},
  459--469\relax
\mciteBstWouldAddEndPuncttrue
\mciteSetBstMidEndSepPunct{\mcitedefaultmidpunct}
{\mcitedefaultendpunct}{\mcitedefaultseppunct}\relax
\EndOfBibitem
\bibitem[Aubin-Tam \emph{et~al.}(2011)Aubin-Tam, Olivares, Sauer, Baker, and
  Lang]{Aubin-Tam_2011}
M.-E. Aubin-Tam, A.~O. Olivares, R.~T. Sauer, T.~A. Baker and M.~J. Lang,
  \emph{Cell}, 2011, \textbf{145}, 257--267\relax
\mciteBstWouldAddEndPuncttrue
\mciteSetBstMidEndSepPunct{\mcitedefaultmidpunct}
{\mcitedefaultendpunct}{\mcitedefaultseppunct}\relax
\EndOfBibitem
\bibitem[Hoang and Cieplak(2000)]{Hoang}
T.~X. Hoang and M.~Cieplak, \emph{J Chem Phys}, 2000, \textbf{112},
  6851--6862\relax
\mciteBstWouldAddEndPuncttrue
\mciteSetBstMidEndSepPunct{\mcitedefaultmidpunct}
{\mcitedefaultendpunct}{\mcitedefaultseppunct}\relax
\EndOfBibitem
\bibitem[Cieplak and Hoang(2003)]{Hoang2}
M.~Cieplak and T.~X. Hoang, \emph{Biophys J}, 2003, \textbf{84}, 475--488\relax
\mciteBstWouldAddEndPuncttrue
\mciteSetBstMidEndSepPunct{\mcitedefaultmidpunct}
{\mcitedefaultendpunct}{\mcitedefaultseppunct}\relax
\EndOfBibitem
\bibitem[Su{\l}kowska and Cieplak(2007)]{JPCM}
J.~I. Su{\l}kowska and M.~Cieplak, \emph{J Phys Condens Matter}, 2007,
  \textbf{19}, 283201\relax
\mciteBstWouldAddEndPuncttrue
\mciteSetBstMidEndSepPunct{\mcitedefaultmidpunct}
{\mcitedefaultendpunct}{\mcitedefaultseppunct}\relax
\EndOfBibitem
\bibitem[Su{\l}kowska and Cieplak(2008)]{models}
J.~I. Su{\l}kowska and M.~Cieplak, \emph{Biophys J}, 2008, \textbf{95},
  3174--3191\relax
\mciteBstWouldAddEndPuncttrue
\mciteSetBstMidEndSepPunct{\mcitedefaultmidpunct}
{\mcitedefaultendpunct}{\mcitedefaultseppunct}\relax
\EndOfBibitem
\bibitem[Sikora \emph{et~al.}(2009)Sikora, Su\l{}kowska, and Cieplak]{plos}
M.~Sikora, J.~I. Su\l{}kowska and M.~Cieplak, \emph{PLoS Comput Biol}, 2009,
  \textbf{5}, e1000547\relax
\mciteBstWouldAddEndPuncttrue
\mciteSetBstMidEndSepPunct{\mcitedefaultmidpunct}
{\mcitedefaultendpunct}{\mcitedefaultseppunct}\relax
\EndOfBibitem
\bibitem[Muthukumar(2007)]{Muthukumar}
M.~Muthukumar, \emph{Annu Rev Biophys Biomol Struct}, 2007, \textbf{36},
  435--450\relax
\mciteBstWouldAddEndPuncttrue
\mciteSetBstMidEndSepPunct{\mcitedefaultmidpunct}
{\mcitedefaultendpunct}{\mcitedefaultseppunct}\relax
\EndOfBibitem
\bibitem[Kirmizialtin \emph{et~al.}(2004)Kirmizialtin, Ganesan, and
  Makarov]{Makarov_2004}
S.~Kirmizialtin, V.~Ganesan and D.~E. Makarov, \emph{J Chem Phys}, 2004,
  \textbf{121}, 10268--10277\relax
\mciteBstWouldAddEndPuncttrue
\mciteSetBstMidEndSepPunct{\mcitedefaultmidpunct}
{\mcitedefaultendpunct}{\mcitedefaultseppunct}\relax
\EndOfBibitem
\bibitem[Huang \emph{et~al.}(2005)Huang, Kirmizialtin, and
  Makarov]{Makarov_2005}
L.~Huang, S.~Kirmizialtin and D.~E. Makarov, \emph{J Chem Phys}, 2005,
  \textbf{123}, 124903\relax
\mciteBstWouldAddEndPuncttrue
\mciteSetBstMidEndSepPunct{\mcitedefaultmidpunct}
{\mcitedefaultendpunct}{\mcitedefaultseppunct}\relax
\EndOfBibitem
\bibitem[West \emph{et~al.}(2006)West, Brockwell, and Paci]{West2006}
D.~K. West, D.~J. Brockwell and E.~Paci, \emph{Biophys J}, 2006, \textbf{91},
  L51--L53\relax
\mciteBstWouldAddEndPuncttrue
\mciteSetBstMidEndSepPunct{\mcitedefaultmidpunct}
{\mcitedefaultendpunct}{\mcitedefaultseppunct}\relax
\EndOfBibitem
\bibitem[Makarov(2009)]{Makarov_2009}
D.~E. Makarov, \emph{Acc Chem Res}, 2009, \textbf{42}, 281--289\relax
\mciteBstWouldAddEndPuncttrue
\mciteSetBstMidEndSepPunct{\mcitedefaultmidpunct}
{\mcitedefaultendpunct}{\mcitedefaultseppunct}\relax
\EndOfBibitem
\bibitem[Szymczak(2013)]{Szymczaktrans}
P.~Szymczak, \emph{Biochem Soc Trans}, 2013, \textbf{41}, 620--624\relax
\mciteBstWouldAddEndPuncttrue
\mciteSetBstMidEndSepPunct{\mcitedefaultmidpunct}
{\mcitedefaultendpunct}{\mcitedefaultseppunct}\relax
\EndOfBibitem
\bibitem[Tian and Andricioaei(2005)]{Tian2005}
P.~Tian and I.~Andricioaei, \emph{J Mol Biol}, 2005, \textbf{350},
  1017--1034\relax
\mciteBstWouldAddEndPuncttrue
\mciteSetBstMidEndSepPunct{\mcitedefaultmidpunct}
{\mcitedefaultendpunct}{\mcitedefaultseppunct}\relax
\EndOfBibitem
\bibitem[Go(1983)]{Go0}
N.~Go, \emph{Annu Rev Biophys Bioeng}, 1983, \textbf{12}, 183--210\relax
\mciteBstWouldAddEndPuncttrue
\mciteSetBstMidEndSepPunct{\mcitedefaultmidpunct}
{\mcitedefaultendpunct}{\mcitedefaultseppunct}\relax
\EndOfBibitem
\bibitem[Takada(1999)]{Takada}
S.~Takada, \emph{Proc Natl Acad Sci U S A}, 1999, \textbf{96},
  11698--11700\relax
\mciteBstWouldAddEndPuncttrue
\mciteSetBstMidEndSepPunct{\mcitedefaultmidpunct}
{\mcitedefaultendpunct}{\mcitedefaultseppunct}\relax
\EndOfBibitem
\bibitem[Wo{\l}ek \emph{et~al.}(2015)Wo{\l}ek, G{\'{o}}mez-Sicilia, and
  Cieplak]{Wolek}
K.~Wo{\l}ek, {\`{A}}.~G{\'{o}}mez-Sicilia and M.~Cieplak, \emph{J Chem Phys},
  2015, \textbf{143}, 243105\relax
\mciteBstWouldAddEndPuncttrue
\mciteSetBstMidEndSepPunct{\mcitedefaultmidpunct}
{\mcitedefaultendpunct}{\mcitedefaultseppunct}\relax
\EndOfBibitem
\bibitem[Galera-Prat \emph{et~al.}(2010)Galera-Prat, G{\'{o}}mez-Sicilia,
  Oberhauser, Cieplak, and Carri{\'{o}}n-V{\'{a}}zquez]{Current}
A.~Galera-Prat, A.~G{\'{o}}mez-Sicilia, A.~F. Oberhauser, M.~Cieplak and
  M.~Carri{\'{o}}n-V{\'{a}}zquez, \emph{Curr Opin Struct Biol}, 2010,
  \textbf{20}, 63--69\relax
\mciteBstWouldAddEndPuncttrue
\mciteSetBstMidEndSepPunct{\mcitedefaultmidpunct}
{\mcitedefaultendpunct}{\mcitedefaultseppunct}\relax
\EndOfBibitem
\bibitem[Eswar \emph{et~al.}(2006)Eswar, Webb, Marti-Renom, Madhusudhan,
  Eramian, Shen, Pieper, and Sali]{modeller}
N.~Eswar, B.~Webb, M.~A. Marti-Renom, M.~S. Madhusudhan, D.~Eramian, M.-Y.~Y.
  Shen, U.~Pieper and A.~Sali, in \emph{Comparative protein structure modeling
  using Modeller.}, John Wiley \& Sons, Inc., 2006, ch.~5\relax
\mciteBstWouldAddEndPuncttrue
\mciteSetBstMidEndSepPunct{\mcitedefaultmidpunct}
{\mcitedefaultendpunct}{\mcitedefaultseppunct}\relax
\EndOfBibitem
\bibitem[Kim \emph{et~al.}(2009)Kim, Chelliah, Kim, Otwinowski, and
  Bezprozvanny]{htt_Q17}
M.~W. Kim, Y.~Chelliah, S.~W. Kim, Z.~Otwinowski and I.~Bezprozvanny,
  \emph{Structure}, 2009, \textbf{17}, 1205--1212\relax
\mciteBstWouldAddEndPuncttrue
\mciteSetBstMidEndSepPunct{\mcitedefaultmidpunct}
{\mcitedefaultendpunct}{\mcitedefaultseppunct}\relax
\EndOfBibitem
\bibitem[Van Der~Spoel \emph{et~al.}(2005)Van Der~Spoel, Lindahl, Hess,
  Groenhof, Mark, and Berendsen]{gromacs}
D.~Van Der~Spoel, E.~Lindahl, B.~Hess, G.~Groenhof, A.~E. Mark and H.~J.~C.
  Berendsen, \emph{J Comput Chem}, 2005, \textbf{26}, 1701--1718\relax
\mciteBstWouldAddEndPuncttrue
\mciteSetBstMidEndSepPunct{\mcitedefaultmidpunct}
{\mcitedefaultendpunct}{\mcitedefaultseppunct}\relax
\EndOfBibitem
\bibitem[Sorin and Pande(2005)]{amber}
E.~J. Sorin and V.~S. Pande, \emph{Biophys J}, 2005, \textbf{88},
  2472--2493\relax
\mciteBstWouldAddEndPuncttrue
\mciteSetBstMidEndSepPunct{\mcitedefaultmidpunct}
{\mcitedefaultendpunct}{\mcitedefaultseppunct}\relax
\EndOfBibitem
\bibitem[Jorgensen \emph{et~al.}(1986)Jorgensen, Chandrasekhar, Buckner, and
  Madura]{TIP3P}
W.~L. Jorgensen, J.~Chandrasekhar, J.~K. Buckner and J.~D. Madura, \emph{Ann N
  Y Acad Sci}, 1986, \textbf{482}, 198--209\relax
\mciteBstWouldAddEndPuncttrue
\mciteSetBstMidEndSepPunct{\mcitedefaultmidpunct}
{\mcitedefaultendpunct}{\mcitedefaultseppunct}\relax
\EndOfBibitem
\bibitem[Humphrey \emph{et~al.}(1996)Humphrey, Dalke, and Schulten]{vmd}
W.~Humphrey, A.~Dalke and K.~Schulten, \emph{J Mol Graphics}, 1996,
  \textbf{14}, 33--38\relax
\mciteBstWouldAddEndPuncttrue
\mciteSetBstMidEndSepPunct{\mcitedefaultmidpunct}
{\mcitedefaultendpunct}{\mcitedefaultseppunct}\relax
\EndOfBibitem
\bibitem[Szymczak and Cieplak(2005)]{clamp}
P.~Szymczak and M.~Cieplak, \emph{Journal of Physics: Condensed Matter}, 2005,
  \textbf{18}, L21\relax
\mciteBstWouldAddEndPuncttrue
\mciteSetBstMidEndSepPunct{\mcitedefaultmidpunct}
{\mcitedefaultendpunct}{\mcitedefaultseppunct}\relax
\EndOfBibitem
\bibitem[Su{\l}kowska \emph{et~al.}(2008)Su{\l}kowska, Su{\l}kowski, Szymczak,
  and Cieplak]{knotjumps}
J.~I. Su{\l}kowska, P.~Su{\l}kowski, P.~Szymczak and M.~Cieplak, \emph{Phys Rev
  Lett}, 2008, \textbf{100}, 058106\relax
\mciteBstWouldAddEndPuncttrue
\mciteSetBstMidEndSepPunct{\mcitedefaultmidpunct}
{\mcitedefaultendpunct}{\mcitedefaultseppunct}\relax
\EndOfBibitem
\bibitem[Sen \emph{et~al.}(2013)Sen, Maillard, Nyquist, Rodriguez-Aliaga,
  Press{\'{e}}, Martin, and Bustamante]{Sen}
M.~Sen, R.~A. Maillard, K.~Nyquist, P.~Rodriguez-Aliaga, S.~Press{\'{e}},
  A.~Martin and C.~Bustamante, \emph{Cell}, 2013, \textbf{155}, 636--646\relax
\mciteBstWouldAddEndPuncttrue
\mciteSetBstMidEndSepPunct{\mcitedefaultmidpunct}
{\mcitedefaultendpunct}{\mcitedefaultseppunct}\relax
\EndOfBibitem
\bibitem[Maurizi and Stan(2013)]{Maurizi}
M.~R. Maurizi and G.~Stan, \emph{Cell}, 2013, \textbf{155}, 502--504\relax
\mciteBstWouldAddEndPuncttrue
\mciteSetBstMidEndSepPunct{\mcitedefaultmidpunct}
{\mcitedefaultendpunct}{\mcitedefaultseppunct}\relax
\EndOfBibitem
\bibitem[Szymczak(2014)]{Szymczakspecial}
P.~Szymczak, \emph{The European Physical Journal Special Topics}, 2014,
  \textbf{223}, 1805--1812\relax
\mciteBstWouldAddEndPuncttrue
\mciteSetBstMidEndSepPunct{\mcitedefaultmidpunct}
{\mcitedefaultendpunct}{\mcitedefaultseppunct}\relax
\EndOfBibitem
\bibitem[Szymczak(2016)]{Szymczakperiodic}
P.~Szymczak, \emph{Sci Rep}, 2016, \textbf{6}, 21702\relax
\mciteBstWouldAddEndPuncttrue
\mciteSetBstMidEndSepPunct{\mcitedefaultmidpunct}
{\mcitedefaultendpunct}{\mcitedefaultseppunct}\relax
\EndOfBibitem
\bibitem[Zatta \emph{et~al.}(2009)Zatta, Drago, Bolognin, and Sensi]{Sensi}
P.~Zatta, D.~Drago, S.~Bolognin and S.~L. Sensi, \emph{Trends Pharmacol Sci},
  2009, \textbf{30}, 346--355\relax
\mciteBstWouldAddEndPuncttrue
\mciteSetBstMidEndSepPunct{\mcitedefaultmidpunct}
{\mcitedefaultendpunct}{\mcitedefaultseppunct}\relax
\EndOfBibitem
\bibitem[Kozlowski \emph{et~al.}(2012)Kozlowski, Luczkowski, Remelli, and
  Valensin]{Kozlowski}
H.~Kozlowski, M.~Luczkowski, M.~Remelli and D.~Valensin, \emph{Coordination
  Chemistry Reviews}, 2012, \textbf{256}, 2129--2141\relax
\mciteBstWouldAddEndPuncttrue
\mciteSetBstMidEndSepPunct{\mcitedefaultmidpunct}
{\mcitedefaultendpunct}{\mcitedefaultseppunct}\relax
\EndOfBibitem
\bibitem[Miller \emph{et~al.}(2012)Miller, Ma, and Nussinov]{Ruth}
Y.~Miller, B.~Ma and R.~Nussinov, \emph{Coordination Chemistry Reviews}, 2012,
  \textbf{256}, 2245--2252\relax
\mciteBstWouldAddEndPuncttrue
\mciteSetBstMidEndSepPunct{\mcitedefaultmidpunct}
{\mcitedefaultendpunct}{\mcitedefaultseppunct}\relax
\EndOfBibitem
\bibitem[Viles(2012)]{Viles}
J.~H. Viles, \emph{Coordination Chemistry Reviews}, 2012, \textbf{256},
  2271--2284\relax
\mciteBstWouldAddEndPuncttrue
\mciteSetBstMidEndSepPunct{\mcitedefaultmidpunct}
{\mcitedefaultendpunct}{\mcitedefaultseppunct}\relax
\EndOfBibitem
\bibitem[Su{\l}kowska \emph{et~al.}(2008)Su{\l}kowska, Sulkowski, Szymczak, and
  Cieplak]{stability}
J.~I. Su{\l}kowska, P.~Sulkowski, P.~Szymczak and M.~Cieplak, \emph{Proc Natl
  Acad Sci USA}, 2008, \textbf{105}, 19714--19719\relax
\mciteBstWouldAddEndPuncttrue
\mciteSetBstMidEndSepPunct{\mcitedefaultmidpunct}
{\mcitedefaultendpunct}{\mcitedefaultseppunct}\relax
\EndOfBibitem
\end{mcitethebibliography}
\providecommand*{\mcitethebibliography}{\thebibliography}
\csname @ifundefined\endcsname{endmcitethebibliography}
{\let\endmcitethebibliography\endthebibliography}{}

\end{document}